\documentclass[aps,prl,twocolumn,superscriptaddress]{revtex4-2}

\usepackage[version=4]{mhchem}
\usepackage{float}
\usepackage{color}

\usepackage{graphicx}% Include figure files
\usepackage{dcolumn}% Align table columns on decimal point
\usepackage{bm}% bold math
\usepackage{hyperref}% add hypertext capabilities
\hypersetup{
    colorlinks=true,
    linkcolor=blue,
    filecolor=magenta,
    urlcolor=red,
    citecolor=blue,
}
\usepackage{glossaries}
\newacronym{CDW}{CDW}{charge density wave}
\newacronym{SDW}{SDW}{spin density wave}
\newacronym{DMRG}{DMRG}{density-matrix renormalization group}
\newacronym{ARPES}{ARPES}{angle-resolved photoemission spectroscopy}

\makeatletter
\renewcommand*{\fnum@figure}{{\normalfont\bfseries \figurename~\thefigure}}  % make bold figure labels
\makeatother

\renewcommand{\figurename}{Fig.}

\begin{document}

\title{Tilted stripes origin in {La$_{1.88}$Sr$_{0.12}$CuO$_{4}$} revealed by anisotropic next-nearest neighbor hopping}

\author{Wei He}
 \email[]{Current address: Department of Condensed Matter Physics and Materials Science, Brookhaven National Laboratory, Upton, New York 11973, USA; whe1@bnl.gov}
 \affiliation{Stanford Institute for Materials and Energy Sciences, SLAC National Accelerator Laboratory, Menlo Park, CA 94025, USA}
 \affiliation{Department of Materials Science and Engineering, Stanford University, Stanford, CA 94305, USA}
\author{Jiajia Wen}
 \affiliation{Stanford Institute for Materials and Energy Sciences, SLAC National Accelerator Laboratory, Menlo Park, CA 94025, USA}
\author{Hong-Chen Jiang}
 \affiliation{Stanford Institute for Materials and Energy Sciences, SLAC National Accelerator Laboratory, Menlo Park, CA 94025, USA}
\author{Guangyong Xu}
 \affiliation{NIST Center for Neutron Research, National Institute of Standards and Technology, Gaithersburg, MD 20899-6102, USA}
\author{Wei Tian}
 \affiliation{Neutron Scattering Division, Oak Ridge National Laboratory, Oak Ridge, Tennessee 37831, USA}
\author{Takanori Taniguchi}
 \affiliation{Institute for Materials Research, Tohoku University, Sendai, 980-8577, Japan}
\author{Yoichi Ikeda}
 \affiliation{Institute for Materials Research, Tohoku University, Sendai, 980-8577, Japan}
\author{Masaki Fujita}
 \affiliation{Institute for Materials Research, Tohoku University, Sendai, 980-8577, Japan}
\author{Young S. Lee}
 \email[]{youngsl@stanford.edu}
 \affiliation{Stanford Institute for Materials and Energy Sciences, SLAC National Accelerator Laboratory, Menlo Park, CA 94025, USA}
 \affiliation{Department of Applied Physics, Stanford University, Stanford, CA 94305, USA}

\date{\today}% It is always \today, today,
             %  but any date may be explicitly specified

\begin{abstract}
Spin- and charge- stripe order has been extensively studied in the superconducting cuprates, among which underdoped \ce{La$_{2-x}$Sr$_{x}$CuO$_{4}$} (LSCO) is an archetype with static spin stripes at low temperatures. An intriguing, but not completely understood, phenomenon in LSCO is that the stripes are tilted away from the high-symmetry Cu-Cu directions. Using high-resolution neutron scattering on LSCO with $x=0.12$, we find two coexisting phases at low temperatures, one with static spin stripes and the other with fluctuating ones, both sharing the same tilt angle. Our numerical calculations using the doped Hubbard model elucidate the tilting’s origin, attributing it to anisotropic next-nearest neighbor hopping $t^{\prime}$, consistent with the material’s slight orthorhombicity. Our results underscore the model’s success in describing specific details of the ground state of this real material and highlight the role of $t^\prime$ in the Hamiltonian, revealing the delicate interplay between stripes and superconductivity across theoretical and experimental contexts.

\end{abstract}

\maketitle

\section{\label{sec:intro}Introduction}
The high-$T_c$ cuprates exhibit complex physical phenomena due to the presence of various phases which may interact with the superconductivity \cite{Fradkin2015}. In recent years, the existence of \gls*{CDW} and \gls*{SDW} order has been observed across many families of cuprates \cite{Keimer2015, Comin2016}. The La-based family is a canonical example where both the spin and charge orders form “stripes” which are especially stable near 1/8 hole doping \cite{Tranquada1995,Tranquada2013,Tranquada2021}. In the stripe model, the doped holes segregate into unidirectional stripes which serve as the antiphase domain boundaries between patches of antiferromagnetically correlated spins. Since the periodicity of the spin order is twice that of the charge order, the wave vectors of these two orders satisfy the relationship $\delta_{charge}=2\delta_{spin}$, which has been confirmed by extensive neutron and x-ray scattering measurements \cite{Tranquada2013,Tranquada2021}.

\begin{figure*}[htb]
 \includegraphics[scale=1]{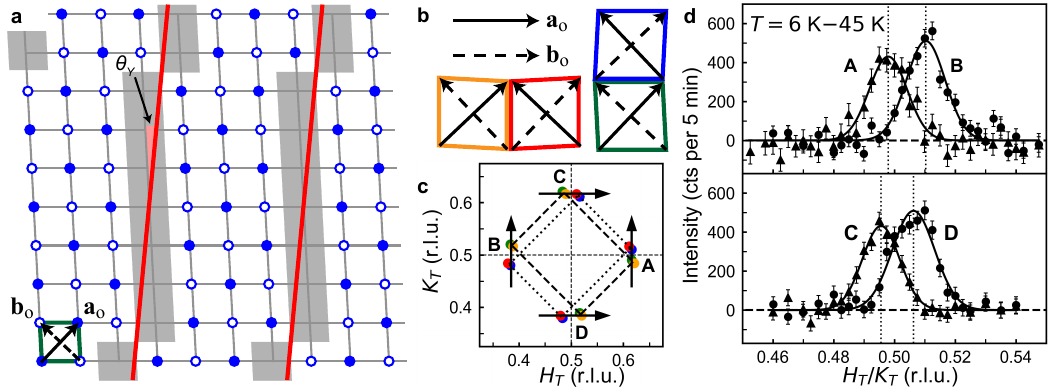}
 \caption{\textbf{Tilted stripes and the crystal structure}. \textbf{a}, A schematic of the CuO$_2$ plane with orthorhombic distortion and the tilted spin- and charge- stripe order. Solid and open blue circles denote spins with opposite directions. The solid green lines show the tetragonal unit cell. The grey regions are the charge stripes served as antiphase domain walls for the magnetic stripes and the red lines show the tilted average stripe direction. \textbf{b}, Tetragonal unit cells of the four structural twin domains in the orthorhombic phase. \textbf{c}, Corresponding possible \gls*{SDW} peaks around (0.5,0.5,0) in reciprocal space. The two rectangles formed by the two sets of peaks (red and blue vs. green and orange) have different centers due to orthorhombic distortion. The solid (dashed) arrows in \textbf{a} and \textbf{b} show the orthorhombic $\mathbf{a_{\rm o}}$ ($\mathbf{b_{\rm o}}$) direction. The orthorhombic distortion and \textit{Y}-shift in \textbf{a}--\textbf{c} are exaggerated for clarity. \textbf{d}, Elastic neutron scattering of the SDW peaks at $6$~K with the $45$~K data subtracted as a background. The solid lines are Gaussian fits to the data. The vertical dotted lines are the fitted peak centers. The trajectory of each scan is denoted in \textbf{c}. Error bars correspond to $\pm \sigma$, where $\sigma$ is the standard deviation.}
 \label{fig:1} 
\end{figure*}

A phenomenon that is not completely understood is that the direction of the stripes is slightly tilted from the underlying Cu-Cu direction. This follows from the observations of small in-plane shifts of both the \gls*{SDW} and \gls*{CDW} peak positions from the high-symmetry directions in orthorhombic cuprates \cite{Lee1999,Kimura2000,Matsushita1999,Fujita2002,Thampy2014,Romer2015,Jacobsen2015,Jacobsen2018,Wen2019,Wen2023Enhanced}. This observation was referred to as the \textit{Y}-shift in early measurements\cite{Kimura2000}, and we will use this term to refer to the observation that denotes tilted stripes. Specifically, in \ce{La$_{2-x}$Sr$_{x}$CuO$_{4}$} (LSCO), the \ce{CuO2} square lattice is deformed upon cooling from a high-temperature tetragonal to a low-temperature orthorhombic structure. As sketched in Fig.~\ref{fig:1}\textbf{a}, the average tilting of the charge domain wall boundaries has a specific orientation with the orthorhombic distortion. This phenomenon was first discovered in oxygen-doped \ce{La_$2$CuO_$4+y$} (LCO) \cite{Lee1999} and subsequently observed in other systems, including LSCO \cite{Kimura2000,Matsushita1999,Jacobsen2015} and \ce{La$_{1.875}$Ba$_{0.125-x}$Sr$_x$CuO$_4$} \cite{Fujita2002}. Recent x-ray scattering of the \gls*{CDW} order in LSCO \cite{Thampy2014,Wen2019,Wen2023Enhanced} also confirmed the existence of the \textit{Y}-shift in the charge stripes, consistent with the \gls*{SDW} order, further corroborating the stripe picture. Such a shift is expected in the phenomenological Landau-Ginzburg model for incommensurate stripe orders in the presence of an orthorhombic distortion \cite{Robertson2006}. However, the microscopic origin of the \textit{Y}-shift was still not clear. 

Recently, new insights regarding the stripe phases and superconductivity have come from numerical simulations using the doped Hubbard model (and $t$-$J$ model) to describe the \ce{CuO$_2$} planes. Half-filled stripes, which are consistent with the periodicities observed for doped LSCO, are found using interaction terms $U$, $t$, and $t'$ \cite{White1998,White1999,Machida1999,Jiang2019Hub,Jiang2020Hub,Chung2020,Lu2024,Xu2024}, where $U$ is the on-site Coulomb repulsion and $t$ ($t'$) is the (next) nearest neighbor hopping. The significance of $t'$ in the cuprates is strongly suggested by the observation that filled stripes (with double the periodicity, inconsistent with LSCO) are favored when only $U$ and $t$ are used in the model \cite{Jiang2020Hub,Qin2020}. Recent \gls*{DMRG} calcuations have shown that the presence of $t'$ is also necessary to induce superconducting correlations \cite{Jiang2019Hub,Jiang2020Hub,Chung2020}. Here, we use \gls*{DMRG} simulations to determine the terms in the Hubbard model that can stabilize the tilted stripes in LSCO with $x=0.12$. We find that $t'$ plays a crucial role, and that the tilted alignment of stripes is highly sensitive to small anisotropies in $t'$, quantitatively consistent with the experimental results.

Another issue to address is whether the tilting phenomenon is particular to static stripes, or if it is generic to the stripe correlations which may also be fluctuating. Spin fluctuations are ubiquitous in the phase diagram of cuprates \cite{Keimer2015}. In the stripe picture, the low-energy spin fluctuations may be thought of as spin waves associated with the ordered stripes, i.e., dynamic stripes \cite{Batista2001,Kivelson2003,Vojta2006,Huang2017}, but the universality of this description is still under debate \cite{Vojta2009}. Due to the broad widths and weak cross sections of the spin fluctuation peaks, it is extremely challenging to detect small peak shifts in the inelastic neutron scattering. Most of the previous studies of the \textit{Y}-shift focused on static stripes, and it remains an open question whether the static and fluctuating spin correlations are characterized by the same stripe tilting.

Here, we present our comprehensive high-resolution neutron scattering study of the \textit{Y}-shift phenomenon on a single crystal of LSCO with $x=0.12$, which has the longest length scale for the spin correlations \cite{Yamada1998}. The sample was mounted in the $(HK0)$ zone, and tetragonal notation is used unless otherwise noted. More details of the experiments can be found in the Methods. First, our elastic scattering results verify the existence of the \textit{Y}-shift in static \gls*{SDW} order. To the best of our knowledge, we determine the spin direction in LSCO for the first time, and provide a new method to determine the interlayer correlations. Then, the inelastic scattering results offer the first evidence for the same \textit{Y}-shift in the dynamic spin stripes, where the spin fluctuation direction is found to be predominantly isotropic. Finally, our numerical calculations using the \gls*{DMRG} method \cite{White1992} explain the microscopic origin of the \textit{Y}-shift. The anisotropy of the next-nearest neighbor hopping term $t^{\prime}$ plays a key role here.

\section{\label{sec:results}Results}
\subsection{\label{sec:results_elastic}Tilted static spin stripes}

\begin{figure*}[htb]
\includegraphics[scale=1]{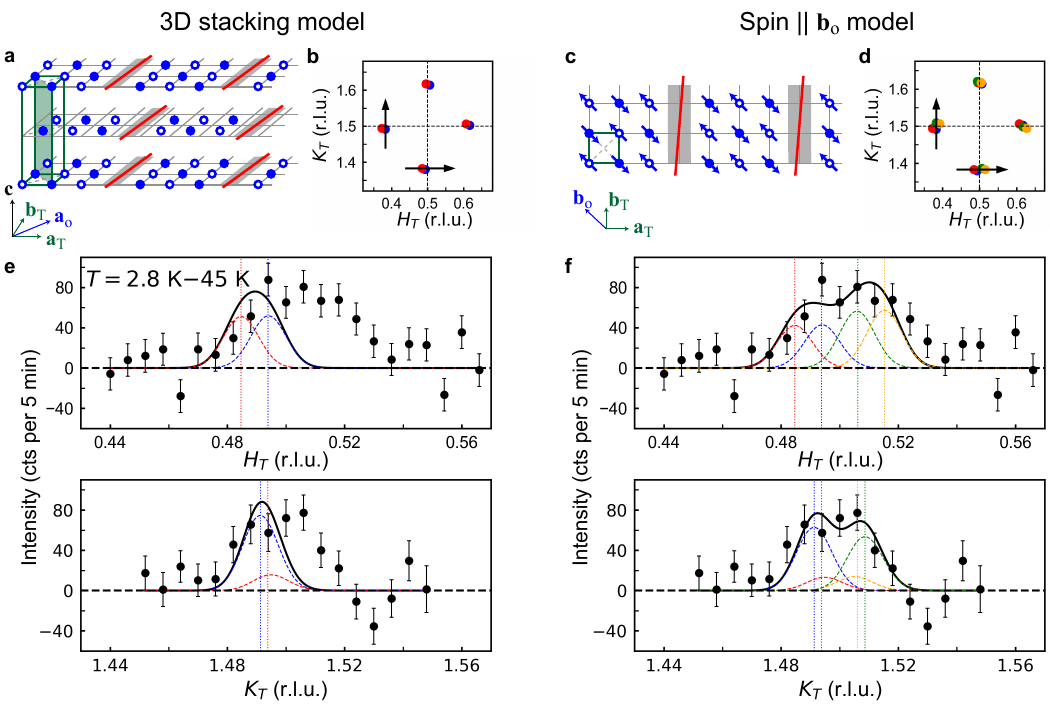}
 \caption{\textbf{Comparison of two models for \gls*{SDW} peaks around (0.5,1.5,0)}. \textbf{a},\textbf{b} and \textbf{e} are for the ``3D stacking model'' while \textbf{c},\textbf{d} and \textbf{f} for the ``Spin $\parallel \mathbf{b}_{\rm o}$ model.'' Panels \textbf{a} and \textbf{c} show schematics of the two models in real space. The solid green lines denote the structural unit cell. The grey regions are the charge stripes and the red lines show the \textit{Y}-shifted effective stripe direction. Solid and open blue circles denote spins with opposite directions. For \textbf{c}, arrows on top of the circles further specify spin directions. The spins in the neighboring layers have fixed relation in the 3D stacking model as shown in \textbf{a}, while this long-range interlayer correlations are absent in the Spin $\parallel \mathbf{b}_{\rm o}$ model. Panels \textbf{b} and \textbf{d} show the corresponding \gls*{SDW} peaks around (0.5,1.5,0) in reciprocal space. The colors of the peaks follow the convention in Fig.~\ref{fig:1}. Panels \textbf{e} and \textbf{f} display elastic neutron scattering of these \gls*{SDW} peaks at $2.8$~K with the $45$~K data subtracted as a background. The solid black lines are fits to the data based on the two models with details in the text. The dashed curve shows the component from each peak with the center at the vertical dotted line. The trajectory of each scan is denoted in \textbf{b} and \textbf{d}. Error bars correspond to $\pm \sigma$, where $\sigma$ is the standard deviation.}
 \label{fig:2}
\end{figure*}

To begin, a proper interpretation of the neutron scattering data requires a careful understanding of the structural twinning in our LSCO single crystal sample. As depicted in Fig.~\ref{fig:1}\textbf{b}, the orthorhombic distortion leads to four possible structural twin domains \cite{Braden1992}. From the precise characterization of nuclear Bragg peaks (see Supplementary Note 1, Supplementary Fig.~1, and Supplementary Tab.~1), we find that our sample has all four domains with similar populations and the orthorhombicity is $0.38(2)^{\circ}$ (calculated as $\frac{b_{\rm o}-a_{\rm o}}{a_{\rm o}}$, where $a_{\rm o}$ and $b_{\rm o}$ are in-plane lattice parameters in orthorhombic notation).

We start with the elastic magnetic scattering of the \gls*{SDW} order near the antiferromagnetic zone center of (0.5,0.5,0). A quartet of \gls*{SDW} peaks can be observed here due to the existence of two magnetic domains (different from the structural twin domains above) with stripe directions perpendicular to each other. Figure~\ref{fig:1}\textbf{d} shows representative scans along the tetragonal $H$ or $K$ directions. The two magnetic domains have roughly the same populations, indicated by the similar intensities between the two pairs of peaks (e.g., peak A vs. C, or peak B vs. D). Within each pair of peaks, a clear shift is observed, corresponding to a \textit{Y}-shift angle of $3.0(2)^{\circ}$, consistent with previous reports \cite{Kimura2000}. However, a mystery observation is the presence of only a single peak in each scan, in contrast to the predicted pattern in Fig.~\ref{fig:1}\textbf{c} based on the existence of four equally-populated structural twin domains. Further inspection (see Supplementary Note 2) indicates that the observed peaks can only originate from the orange and green structural domains, both of which have the shorter orthorhombic $\mathbf{a}_{\rm o}$ axis nearly aligned with (0.5,0.5,0). The in-plane correlation length is determined to be $123(7)$~\AA{} (see Supplementary Fig.~2 and Supplementary Tab.~2).

We propose two models to explain the missing contributions from the other two twin domains. As illustrated in Fig.~\ref{fig:2}\textbf{a}, the first model involves the coupling between layers and therefore is named the ``3D stacking model''. The depicted stacking arrangement (which is locally similar to that in \ce{La$_2$CuO$_4$} \cite{Vaknin1987}) results in nearly zero intensity around (0.5,0.5,0) for the red and blue domains. The second model depends on the spin direction, relying on the fact that the magnetic neutron scattering cross-section is only sensitive to components of the spin $\mathbf{S}$ that are perpendicular to the wave vector $\mathbf{Q}$. Hence, if the spins are pointing along the orthorhombic $\mathbf{b}_{\rm o}$ direction, as displayed in Fig.~\ref{fig:2}\textbf{c}, the red and blue domains will give nearly zero intensities around (0.5,0.5,0) regardless of the stacking arrangement between layers. Quantitatively, considering that the intensities from the red and blue domains around (0.5,0.5,0) are less than $\sim5\%$ of those from the green and orange domains in Fig.~\ref{fig:1}\textbf{d}, the spins need to be either correlated over at least 14.3~\AA{} in the $L$ direction using a finite-size domain analysis (see Supplementary Note 2D) \cite{x-ray} in the first model, or fixed to the orthorhombic $\mathbf{b}_{\rm o}$ direction within $10^{\circ}$ in the second model. 

Both spin models proposed above are reminiscent of the spin structure of the parent compound \ce{La$_2$CuO$_4$} \cite{Vaknin1987} and in oxygen-doped LCO \cite{Lee1999}. To further distinguish the two models, one approach is to measure the $L$-dependence of the \gls*{SDW} intensity as in Ref.~\cite{Lee1999}. However, this method involves switching sample scattering geometry. As an alternate approach, we implement a strategy to distinguish these two models by going to higher Brillouin zones within the same $(HK0)$ geometry. As shown in Fig.~\ref{fig:2}\textbf{d} near the (0.5,1.5,0) position, since the wave vector $\mathbf{Q}$ no longer coincides with the spin directions, peaks from four domains should all contribute to the neutron scattering. However, if the 3D stacking model is correct, then peaks from the orange and green domains would be forbidden as drawn in Fig.~\ref{fig:2}\textbf{b}. Figure~\ref{fig:2}\textbf{e} demonstrates that the measured peak profiles clearly deviate from the expectations for the 3D stacking model. Therefore, the second model with spins aligned along the orthorhombic $\mathbf{b}_{\rm o}$-axis should be preferred. Indeed, the fits based on this ``spin $\parallel \mathbf{b}_{\rm o}$ model'' gives significantly improved agreement as shown in Fig.~\ref{fig:2}\textbf{f}. Moreover, by co-fitting with the \gls*{SDW} peaks near (0.5,0.5,0), we can further quantify the interlayer correlation length and the volume fractions of stripe ordered phases (see Supplementary Note 2). Using a finite-size domain analysis \cite{x-ray}, the interlayer correlation length is determined to be $7(1)$~\AA{}, corresponding to a slight modulation of intensities by $(20\pm15)\%$ along $L$.

The intensities of the fits in Fig.~\ref{fig:2}\textbf{f} (colored lines) indicate the static \gls*{SDW} does not occur homogeneously throughout the sample. Rather, only a partial fraction of the sample contains static \gls*{SDW}, where the red and blue structural domains contain $2.9(6)$ times more of the static \gls*{SDW} phase volume fraction compared to the other two domains, as evidenced in Supplementary Fig.~3. By normalizing the \gls*{SDW} peak intensities with the nuclear Bragg peaks, the average ordered moment at $6$~K is deduced to be at most $0.07(1)~\mu{_B}$, close to the previous reported values in LSCO \cite{Kimura1999a,Kimura1999} but only about half of that in oxygen-doped LCO \cite{Lee1999}. Assuming a local moment value of $0.36~\mu{_B}$ from a previous $\mu$SR report \cite{Savici2002}, this suggests the magnetic volume fraction is less than $\sim5\%$. This value is lower than the ordered fraction of $18\%$ from the $\mu$SR study \cite{Savici2002}, possibly due to sample variation and differing sensitivities of the probes. $\mu$SR is a more localized probe, whereas neutron scattering detects long-range ordered spins. The implications of the minority static \gls*{SDW} phase are further discussed in the Discussion Section.

\subsection{\label{sec:results_inelastic}Determination of tilting for the fluctuating spin stripes}

\begin{figure}[h!]
 \includegraphics[scale=1]{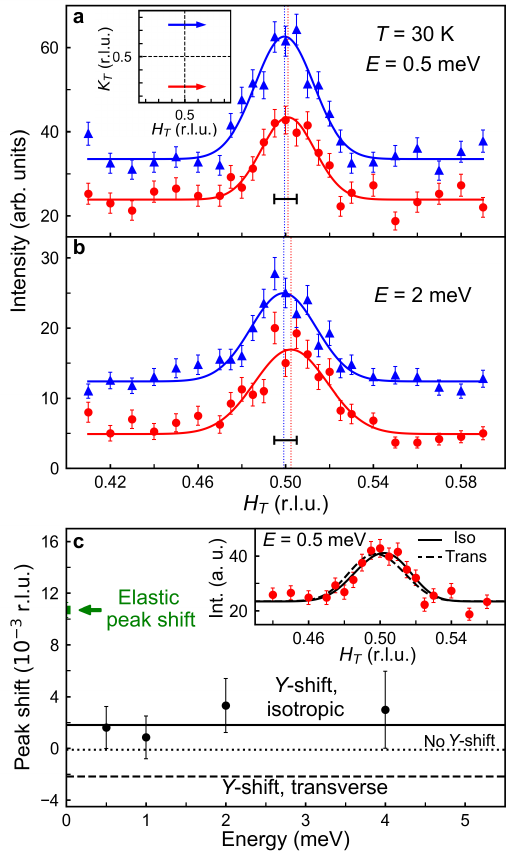}%
 \caption{\textbf{Energy dependence of the dynamic spin stripes}. \textbf{a},\textbf{b}, Inelastic neutron scattering of the spin stripes at $30$~K for an energy transfer of $E = 0.5$~meV (\textbf{a}) and $E = 2$~meV (\textbf{b}). The solid lines are Gaussian fits with a constant background and the vertical dotted lines are the fitted peak centers. The trajectory of each scan is denoted in the inset of \textbf{a}. The horizontal bars represent the elastic peak shift. The blue data are shifted vertically for clarity. \textbf{c}, Energy dependence of the peak shift between the fitted centers of a pair of peaks as shown in \textbf{a},\textbf{b}, and Supplementary Fig.~5. Solid and dashed lines are the expected peak shift for isotropic and transverse spin fluctuations $\Delta \mathbf{S}$ with \textit{Y}-shift, respectively, while the dotted line for isotropic spin fluctuations without \textit{Y}-shift. Inset: the same red data in \textbf{a} with fits assuming isotropic (solid line) and transverse (dashed line) excitations with \textit{Y}-shift. Error bars correspond to $\pm \sigma$, where $\sigma$ is the standard deviation.}
 \label{fig:3}
\end{figure}

\begin{figure}[tbp]
 \includegraphics[scale=1]{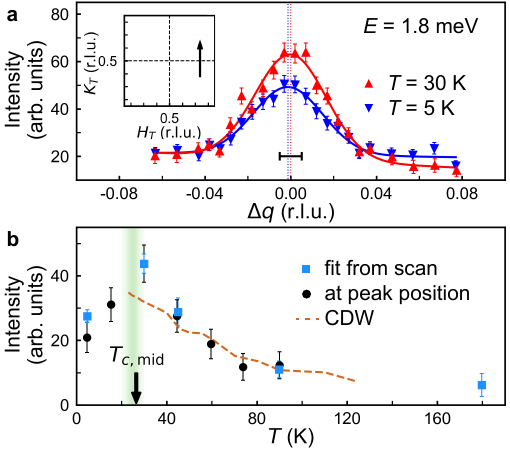}
 \caption{\textbf{Temperature dependence of the dynamic spin stripes}. \textbf{a}, Representative scans at $5$~K and $30$~K for an energy transfer of $E = 1.8$~meV. The scan trajectory is shown in the inset. The solid lines are Gaussian fits with a linear background and the vertical dotted lines are the fitted peak centers. The horizontal bar represents the elastic peak shift. \textbf{b}, Integrated intensity extracted from the Gaussian fits (blue squares). The black circles are intensities from one-point measurement at the peak position and scaled to match the fitted integrated intensity at $30$~K. The arrow indicates the midpoint temperature ($26.5$~K) of the superconducting transition $T_{c\rm{, mid}}$ \cite{Kimura2000}. The diffuse green vertical line indicates the \gls*{SDW} onset temperature varying from $\sim20$~K ($\mu$SR \cite{Savici2002}) to $\sim30$~K (neutron \cite{Kimura1999a}). The brown dashed line is the integrated intensity of \gls*{CDW} peak measured by resonant soft x-ray scattering \cite{Wen2019} and scaled to match our data. Error bars correspond to $\pm \sigma$, where $\sigma$ is the standard deviation.}
 \label{fig:4}
\end{figure}

Armed with the detailed knowledge learned from elastic neutron scattering (such as the structural domain population, magnetic structure, \textit{Y}-shift angle in the static \gls*{SDW} phase, etc.), we next turn to high-resolution inelastic scattering to study the low-energy excitations of the spin stripes. Figure~\ref{fig:3}\textbf{a}--\textbf{b} shows scans around (0.5,0.5,0) at two representative energy transfers at $30$~K, which is close to the onset temperature of the SDW order $T_{\rm sdw}$ \cite{Kimura1999a} and has the highest intensities (evidenced in Fig.~\ref{fig:4}\textbf{b}. The energy dependence of the inelastic peak shift is plotted in Fig.~\ref{fig:3}\textbf{c}. Although the shifts here are much smaller than the elastic case, their consistently non-zero values indicate that the dynamic stripes also have a \textit{Y}-shift. Note that the case assuming no \textit{Y}-shift for the inelastic scattering is given by the dotted line near zero shift (this value includes a very small correction taking into account the structural domain populations). Our data show that the fluctuating stripes are tilted with the same tilt angle of about 3.0$^\circ$ as observed for the static spin stripes. That is, our analysis indicates that the inelastic peaks appearing at the same positions as their elastic counterparts in the low-energy limit, consistent with the Goldstone theorem. Recent findings that the \textit{Y}-shift exists in the \gls*{CDW} order at $\sim T_{\rm sdw}$ in LSCO \cite{Thampy2014,Wen2019,Wen2023Enhanced} with the shift angle similar to low-temperatures are consistent with this notion.

The reason that the apparent \textit{Y}-shift is much smaller than that in the elastic scattering is simply an averaging effect of the structural twin domains. As seen in the analysis of the elastic magnetic cross-section, the inelastic scattering intensities are also affected by the neutron polarization factor and interlayer correlations. Specifically, for the inelastic peaks around (0.5,0.5,0), the interlayer correlations reduce the intensities of the red and blue domains by $\sim20\%$. The neutron polarization factor has a bigger effect, resulting in essentially zero intensities for the longitudinal spin fluctuations $\Delta \mathbf{S}$ from the red and blue domains and essentially zero intensities for the in-plane transverse spin fluctuations of the orange and green domains. Here, longitudinal and transverse spin fluctuation directions are with respect to the static spin direction. With these constraints in mind, we calculate the weighted peak position for various cases.

We can first rule out the scenario in which the inelastic signal arises only from the magnetically ordered regions. In that case, the peaks have dominant contributions from the red and blue domains due to their larger stripe volume fraction and therefore shift to the opposite direction (see Supplementary Fig.~6). Hence, we can conclude that the inelastic scattering likely originates from the whole sample rather than just the magnetically ordered regions. Then we test two extreme cases: purely isotropic excitations and purely transverse excitations. As shown in Fig.~\ref{fig:3}\textbf{c} and Supplementary Fig.~8, the isotropic model clearly describes the data better. If we adopt this isotropic model, the fitted intrinsic tilt angle is $3.1(9)^{\circ}$, which, within the errors, is the same as that for the static spin stripes ($3.0(2)^{\circ}$). The observed small positive shift is primarily caused by the slight decrease of the intensities from the red and blue domains due to the finite interlayer correlations. The observed difference in the peak positions for the elastic and inelastic scattering is simply a domain-averaging effect, and phase separation explains the existence of the isotropic spin excitations (see further discussions in the Discussion Section).

We also investigate the temperature dependence of the inelastic scattering (see Fig.~\ref{fig:4} and Supplementary Fig.~9). Even at $5$~K, which is well below $T_{\rm sdw}$, no obvious change of the peak position is observed relative to the $30$~K data. Therefore, we can extend our previous conclusion of isotropic excitations to the low temperature regime. This interpretation is consistent with recent results in oxygen-doped LCO \cite{Tutueanu2021}. Another type of phase separation scenario has been suggested in LSCO emphasizing a dip in the excitation spectrum around $\sim4$~meV below $T_c$ as a hidden spin gap for the superconducting regions.\cite{Kofu2009} We do not observe a shift of the inelastic peak positions for energy transfers below $4$~meV upon cooling, as shown in Fig.~\ref{fig:4}\textbf{a}. This could be due to a distribution of gap sizes and the energy we choose here (1.8~meV) is still above the spin gap in the majority of the sample. Energy-dependent measurements in the ordered state would be crucial to investigate this hidden-spin-gap hypothesis.

We also note that in Fig.~\ref{fig:4}\textbf{b} the intensities of spin fluctuations above $T_{\rm sdw}$ closely follow the trend of the \gls*{CDW} intensities recently measured using resonant soft x-ray scattering \cite{Wen2019}. The extracted dynamical spin correlation length ($62(10)$~\AA{} at $E=0.5$~meV, see Supplementary Note 3 and Supplementary Fig.~7) is close to the \gls*{CDW} correlation length \cite{Thampy2014, Wen2019, Wen2023Enhanced}. 

\subsection{\label{sec:results_dmrg}Numerical calculations: understanding the origin of the tilting}

\begin{figure*}[htbp]
 \includegraphics[scale=1]{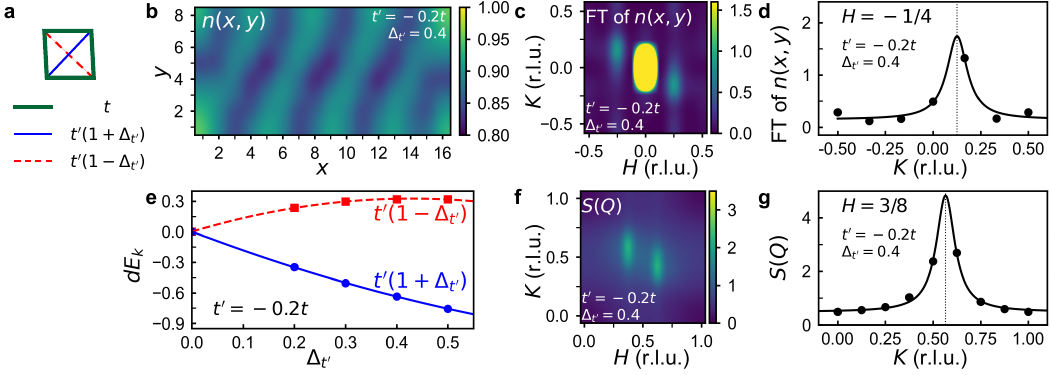}
 \caption{\textbf{Numerical simulations of the origin of the stripe tilting}. \textbf{a}, The unit cell with orthorhombic distortion to illustrate the electron hopping terms used in the calculations. \textbf{b}, The electron density distribution $n(x,y)$ calculated for $t'=-0.2t$ and $\Delta_{t'}=0.4$. \textbf{c}, The Fourier transform of the electron density distribution in \textbf{b}. \textbf{d}, $K$-cut of the charge ordering peaks at $H=-0.25$ in \textbf{c}. \textbf{e}, The kinetic energy differences as defined in the main text for $t'=-0.2t$. The lines are quadratic fits to the data. \textbf{f}, The static spin structure factor $S(\mathbf{Q})$ calculated for $t'=-0.2t$ and $\Delta_{t'}=0.4$. \textbf{g}, $K$-cut of the spin ordering peaks in $S(\mathbf{Q}$) at $H=0.375$. The maps in \textbf{b}, \textbf{c}, and \textbf{f} are produced using multiquadric interpolation method \cite{matplotlib}. The solid lines in \textbf{d} and \textbf{g} are Lorentzian fits with a constant background to determine the peak center (denoted by the vertical dotted lines). The Lorentzian peak widths for charge ordering and spin ordering peaks in the fitting are fixed to the best overall value based on data with all different $\Delta_{t'}$, respectively.}
 \label{fig:5}
\end{figure*}

\begin{figure}[htbp]
 \includegraphics[scale=1]{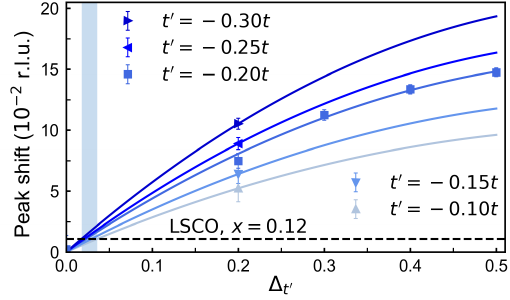}
 \caption{\textbf{The spin stripe tilting as a function of $\Delta_{t'}$ for different $t'$}. The tilting is measured by the peak shift along $K$ direction between the pair of peaks in $S(\mathbf{Q})$ maps (shown in Fig.~\ref{fig:5}\textbf{f} and Supplementary Figs.~11 and 14). The solid line for $t'=-0.2t$ is a quadratic fit to the data. Other lines are the same quadratic curve as for $t'=-0.2t$ but scaled to match the corresponding data. The horizontal dashed line indicates the elastic peak shift measured in our LSCO sample. The vertical bar represents the range of our estimated $\Delta_{t'}$ determined by the intersection of the fitted curves and the dashed line. Error bars correspond to $\pm \sigma$, where $\sigma$ is the standard deviation.}
 \label{fig:6}
\end{figure}

The experimental observation that the same tilting arrangement is inherent to both static and dynamic stripes calls for a microscopic explanation. Here, we study the spatially anisotropic $t$-$t'$ Hubbard model on the square lattice using the \gls*{DMRG} method \cite{White1992}, which is defined by the Hamiltonian%
\begin{eqnarray}\label{Eq:Ham}
%H=-\sum_{\langle ij\rangle\sigma} t_{ij}
H=-\sum_{ij\sigma} t_{ij}
\left(\hat{c}^\dagger_{i\sigma} \hat{c}_{j\sigma} + h.c.\right) + U\sum_i\hat{n}_{i\uparrow}\hat{n}_{i\downarrow}.
\end{eqnarray}
This is one of the simplest models that respect the orthorhombic symmetry in real materials. Here $\hat{c}^\dagger_{i\sigma}$ ($\hat{c}_{i\sigma}$) is the electron creation (annihilation) operator on site $i=(x_i,y_i)$ with spin-$\sigma$, and $\hat{n}_{i\sigma}$ is the electron number operator. The electron hopping amplitude $t_{ij}=t$ if $i$ and $j$ are nearest neighbors and $t_{ij}=t'(1\pm \Delta_{t'})$ for next-nearest neighbors where $\Delta_{t'}$ is the spatial anisotropy related to the orthorhombic distortion as shown in Fig.~\ref{fig:5}\textbf{a}. $U$ is the on-site Coulomb repulsion. We set $t=1$ as an energy unit with interaction $U=12t$ and report results for $t'=-0.3t\sim -0.1t$, which corresponds to the typical values reported in the literature for LSCO \cite{Pavarini2001,Damascelli2003,Peng2017} with doping near $\delta=12.5\%$, the hole concentration used in our study. The main results are shown in Figs.~\ref{fig:5} and \ref{fig:6} and the details are given below. Additional details can be found in Supplementary Note 4 and Supplementary Figs.~10--20.

We first calculate the electron density distribution $n(x,y)=\langle \hat{n}(x,y)\rangle$, where an example is shown in Fig.~\ref{fig:5}\textbf{b} for $t'=-0.2t$ and $\Delta_{t'}=0.4$. In the absence of spatial anisotropy, i.e., $\Delta_{t'}=0$, we find the charge stripe (parallel to the $y$ direction) of wavelength $\lambda_c\approx 4$ (see Supplementary Fig.~10\textbf{a}, consistent with the half-filled charge stripes observed in previous studies \cite{White1998,White1999,Jiang2019Hub,Jiang2020Hub}. In our simulations, the charge stripe being bond-centered is due to the even number of sites along the $x$ direction in the calculation. Our results show that the charge stripe is tilted when $\Delta_{t'}>0$ as shown in Fig.~\ref{fig:5}\textbf{b}--\textbf{d}, consistent with the experimental observations. This tilting reduces the kinetic energy of the charge stripe.\cite{Zaanen1989,Kato1990} To support this, we further calculate the kinetic energy for both $t'(1\pm \Delta_{t'})$ bonds as $E_k(\pm \Delta_{t'})=-t'(1\pm \Delta_{t'}) \sum _{ij \in (1\pm \Delta_{t'})\rm{-bonds}}\langle c_{i\sigma}^\dagger c_{j\sigma}\rangle$. The kinetic energy difference $dE_k(\pm \Delta_{t'})=E_k(\pm \Delta_{t'})-E_k(\Delta_{t'}=0)$ is shown in Fig.~\ref{fig:5}\textbf{e}. While the kinetic energy along the $(1-\Delta_{t'})$-bonds is increased, i.e., $dE_k(-\Delta_{t'})>0$, the kinetic energy along the $(1+\Delta_{t'})$-bonds is reduced, so that the total kinetic energy $dE_k=dE_k(\Delta_{t'})+dE_k(-\Delta_{t'})<0$ when $\Delta_{t'}>0$. We note that a finite value of $\Delta_{t'}>0.1$ is necessary to give observable tilting of the stripes due to limited sample size along $y$ direction in the calculations, although the eight-leg ladder size in \gls*{DMRG} calculations with the presented accuracy is already unprecedented as far as we are aware.

To describe the magnetic properties of the system, we have also calculated the static spin structure factor, which is defined as%
\begin{eqnarray}
S(\mathbf{Q})=\frac{1}{N}\sum_{i,j=1}^N e^{i\mathbf{k}\cdot (\mathbf{r}_i - \mathbf{r}_j)} \langle\mathbf{S}_i\cdot \mathbf{S}_j\rangle. \label{Eq:Sq}
\end{eqnarray}
Here the wavevector $\mathbf{Q}=(k_x,k_y)=k_x\mathbf{b}_x + k_y\mathbf{b}_y$ in reciprocal space is defined by the reciprocal vectors $\mathbf{b}_x=(2\pi,0)$ and $\mathbf{b}_y=(0,2\pi)$. Consistent with previous studies \cite{White1998,White1999,Jiang2019Hub,Jiang2020Hub}, we find that the spin-spin correlations show a spatial oscillation of wavelength $\lambda_s$ that is mutually commensurate with the \gls*{CDW} order, i.e., $\lambda_s\approx 2\lambda_c$. As a result, $S(\mathbf{Q})$ is peaked at the momentum $(k_x,k_y)$ where $k_x=0.5\pm 0.125$ and $k_y=0.5$. Similar with the charge stripe, we find that the peak position of $S(\mathbf{Q})$ is shifted with $\Delta_{t'}$ as shown in Fig.~\ref{fig:5}\textbf{f} and \textbf{g}. The peak shift, i.e., the difference of peak positions along $y$ direction for this pair of peaks in $S(\mathbf{Q})$, and its dependence on $\Delta_{t'}$ is given in Fig.~\ref{fig:6}. We focus on $t'=-0.2t$, and find that the relationship between the peak shift and $\Delta_{t'}$ can be well fitted by a quadratic function, which is commonly used in the analysis of \gls*{DMRG} data \cite{Jiang2019Hub,Qin2020}. The use of the quadratic function passing origin without a threshold value in $\Delta_{t'}$ is also consistent with the previous phenomenological study \cite{Robertson2006}, which showed that the tilting is proportional to the orthorhombicity to first order approximation.

Comparing our results with experimental observation allows us to estimate the spatial anisotropy of $t'$ in the real material. If we interpolate the above empirical curve to the experimental observed value of the peak shift in our LSCO sample, the expected anisotropy is $\Delta_{t'}=2.3(1)\%$. This estimation of $\Delta_{t'}$ depends somewhat on the value of $t'$. For $t'=-0.3t\sim -0.1t$, if we assume the peak shift follows the same quadratic form, the estimated $\Delta_{t'}$ can vary between $1.8\% \sim 3.6\%$, which is pretty close to the prediction ($\Delta_{t'}\gtrsim1.5\%$) from previous studies \cite{Yamase2000, Yamase2003}. Note that $\Delta_{t'}$ has not been determined experimentally yet, even with advanced spectroscopic techniques such as \gls*{ARPES}. This could be due to its small effect on the band structure. Additionally, the presence of the structural twinning may further obscure the interpretation of such measurements.

We can estimate the energy scale of the tilted stripes. The difference between the hopping integral for the two diagonal directions is $|2t'\Delta_{t'}|$. Taking $\Delta_{t'} = 0.023$, $t'=-0.2t$, and $t=0.43$~eV \cite{Hybertsen1990,Pavarini2001}, we have $|2t'\Delta_{t'}|= 4.0$~meV$\sim46$~K. This is consistent with the temperature range where the \textit{Y}-shift can be clearly detected with neutron scattering. At higher temperatures, thermal broadening may obscure the small shift of the peak positions.

The above calculations are performed on a system size of $N=16\times 8$, where $L_x=16$ and $L_y=8$ are the number of sites in $\hat{x}$ and $\hat{y}$ directions, respectively. The reason that we use eight-leg ladder system size is because this appears to be the minimal size required to produce half-filled stripes with ordinary $d$-wave symmetry compatible with experimental observations. For comparison, the ground state superconducting pairing symmetry is the plaquette $d$-wave on the hole doped side ($t'<0$) for the four-leg ladder system \cite{Chung2020}, while half-filled charge stripes are forbidden for the six-leg ladder system if superconducting Cooper pairs are present \cite{Jiang2024}. To show that the boundary effects and finite-size effects have negligible impacts on our fitted peak shifts, we performed more numerical calculations with different system sizes ($N=12\times 8$ and $N=8\times 8$). The results indeed show good convergence between $N=12\times 8$ and $N=16\times 8$ system sizes, supporting the accuracy of our calculations. More details can be found in Supplementary Figs.~17--20. We note that our results are based on local charge and spin modulations instead of the long-range decaying behavior of the correlation functions and therefore should not be affected by the exact nature of the ground state being stripes or $d$-wave superconductivity in the model.

\section{\label{sec:discussion}Discussion}
The primary conclusions from our combined experimental and numerical work can be summarized in three main points: 1) two types of phases coexist in our sample—a minority phase with static \gls*{SDW} and a majority phase with fluctuating spin stripes, 2) both phases have tilted stripes (\textit{Y}-shift) with the same degree of tilting, and 3) the origin of the tilting can be explained by a small anisotropy in the hopping interaction $t'$. We elaborate on the implications of these results below.

Our elastic and inelastic neutron scattering results are naturally explained by the presence of two distinct phases (one with static spin stripes and one with fluctuating spin stripes). As discussed in the elastic scattering results, we estimate that only a small fraction ($\sim5\%$) of the sample is magnetically ordered. The observation of a partially magnetically ordered phase in LSCO has been previously reported by $\mu$SR studies \cite{Savici2002, Chang2008}. This is further supported by recent x-ray scattering studies that two types of \gls*{CDW} orders coexist in LSCO, where the longer-range ordered \gls*{CDW} takes place in a minority fraction of the sample.\cite{Wen2019,Wen2023Enhanced} This suggests that the phase with static \gls*{SDW} has a type of \gls*{CDW} order which is distinguished by having longer range correlations.

The low symmetry of the \textit{Y}-shift positions combined with high-resolution measurements enabled us to determine the spin direction and interlayer correlations. So far as we know, this is the first time that the spin direction is explicitly determined to be along the orthorhombic $\mathbf{b}_{\rm o}$ direction in LSCO. The rather short interlayer correlation length of $\sim 7$~\AA{} is consistent with the previous reports in both \gls*{SDW} and \gls*{CDW} studies in LSCO \cite{Lake2005,Romer2015,Croft2014,Christensen2014}, affirming the 2D nature of the stripe order. Such short interlayer correlations could be explained by stacking faults between layers (see Supplementary Fig.~4). The spin direction and local interlayer correlations are similar to those in the parent compound \ce{La$_2$CuO$_4$}. This reiterates that doped antiferromagnets are essential to the description of high-$T_c$ cuprates \cite{Orenstein2000}.

The inelastic scattering measurements of the isotropic nature of the spin fluctuations requires the presence of an additional phase. Since spin-wave excitations can only contain transverse fluctuations below $T_{\rm sdw}$, the measured isotropic excitations indicates that there is a majority phase that is not magnetically ordered, characterized with isotropic spin excitations. In contrast, the minority phase with static \gls*{SDW} order should have only transverse spin excitations. Therefore, the peak shift in Fig.~\ref{fig:3}\textbf{c} represents the average of the isotropic and transverse contributions, weighted by their respective volume fractions, which is dominated by the non-magnetically ordered phase. This majority phase should also be responsible for the superconductivity in the sample, which is a bulk superconductor. The minority phase may not be superconducting due to competition with the static \gls*{SDW}.

As discussed in the inelastic scattering results, the intensities and widths of the fluctuating spin stripe peaks has a close correlation to the \gls*{CDW} peaks observed with x-rays. These intimate relationships suggest that the two signals detected by different tools probably originate from the same majority phase. This is consistent with the recent NMR finding that increasing charge ordered domains trigger the glassy freezing of Cu spins below $T_{\rm cdw}$ \cite{Arsenault2020}.

We find that both phases contain tilted stripe correlations with the same tilt angle of about 3.0$^\circ$. Our data and analysis indicate that the same \textit{Y}-shift is an intrinsic property of both static and fluctuation phases. This contrasts with a previous study of oxygen-doped LCO where it was concluded that the spin fluctuations have a significantly different stripe tilt angle compared to the static \gls*{SDW} order.\cite{Jacobsen2018} Our numerical results which explain the \textit{Y}-shift indicate that the static and fluctuating stripes should have the same tilting, as it depends only on the underlying orthorhombicity. 

\gls*{DMRG} calculations provide a direct and unbiased way to understand the origin of the \textit{Y}-shift phenomenon. The parameters in our state-of-the-art calculations on eight-leg ladders are consistent with the structure of real materials. The anisotropy of $t'$ may arise from inequivalent \ce{CuO_6} octahedron tilts and unequal lattice parameters between the orthorhombic $a_{\rm o}$ and $b_{\rm o}$ axes. Here, it is estimated to be $\gtrsim1.5\%$ in LSCO with $x=0.12$ \cite{Yamase2000, Yamase2003}. Our \gls*{DMRG} prediction for the anisotropy $\Delta_{t'} = 1.8\% \sim 3.6\%$ matches this expected value for LSCO very well, and much better than the mean field approach based on Fermi surface nesting \cite{Yamase2000, Yamase2003}. Moreover, our results reveal the importance of kinetic energy considerations in stripe formation.\cite{Zachar1998} Microscopically, the anisotropy of $t'$ causes the stripes to tilt towards the short $\mathbf{a}_{\rm o}$ axis, the direction with larger hopping magnitude and gain in kinetic energy. This is an obvious manifestation of the role of kinetic energy term in stabilizing charge stripes by having delocalized holes along the stripes \cite{Zaanen1989,Emery1999}. Our \gls*{DMRG} results also qualitatively agree with an early exact diagonalization study \cite{Vos2003}, which found that an anisotropic $t'$ in the extended $t$-$J$ model results in a preferred orientation of stripes formed by doped holes.

The explanatory power of our results can further be tested by their applicability to similar orthorhombic cuprate materials. Indeed, the \textit{Y}-shift has been reported in both \gls*{CDW} and \gls*{SDW} orders with various dopants in La-based cuprates \cite{Lee1999,Kimura2000,Matsushita1999,Fujita2002,Thampy2014,Romer2015,Wen2019,Wen2023Enhanced}. Tilted \gls*{SDW} order was also found in LSCO at lower doping level of x=0.07 \cite{Jacobsen2015}, with a surprisingly large tilt angle. In the above cases, the larger tilt angle goes hand-in-hand with a larger structural orthorhombicity, as expected in our explanation based on anisotropic $t'$. A different class of stripes (i.e., diagonal stripes) have been observed in lightly-doped LSCO \cite{Wakimoto1999,Fujita2002b,Matsuda2002} and \ce{La$_{2-x}$Ba$_{x}$CuO$_{4}$} (LBCO) \cite{Dunsiger2008} below $x\sim0.055$, coincident with a superconductor-to-insulator transition. Since these diagonal stripes are aligned along the shorter $\mathbf{a}_{\rm o}$ direction, this could be regarded as the extreme case of our tilted stripe explanation. Such unidirectional diagonal stripes are not limited to cuprates—they were also found in the insulating nickelates \cite{Hucker2006}, implying the possible generality of this mechanism for the selection of the stripe direction in systems with similar orthorhombic symmetry.

Uniaxial strain engineering offers a powerful tuning knob to artificially modify the hopping strengths along specific directions. We note that a recent x-ray scattering study \cite{Wang2022} observed a decrease in stripe tilt angle in LSCO under uniaxial pressure applied along one of the tetragonal directions. This agrees with our kinetic energy argument that the stripe direction should be sensitive to the anisotropy of the hopping terms. Such uniaxial pressure can result in an anisotropy in $t$. In another recent elastic neutron scattering experiment \cite{Simutis2022}, uniaxial pressure was used to select a single magnetic domain in LSCO with the stripe direction aligning along the larger $t$ direction. LBCO has intrinsic anisotropy in $t$ due to octahedron rotation along the Cu-Cu bond directions. A recent uniaxial strain experiment in LBCO reports surprisingly different behaviors in elastic and inelastic channels in terms of incommensurability \cite{Kamminga2023}. In the future, it would be appealing to investigate numerically how the tilt angle would change with both anisotropic $t$ and $t'$ terms in the Hamiltonian.

These results show an excellent match between \gls*{DMRG} calculations for the doped Hubbard model and the specific ground state details observed in \ce{La$_{1.88}$Sr$_{0.12}$CuO$_{4}$} --- attesting to the appropriateness of the model for the cuprates. Our results further stress the importance of $t'$ in the Hamiltonian. The presence of $t'$ is known to stabilize half-filled stripes and superconductivity, which are both present in our LSCO sample.\cite{White1999,Jiang2019Hub,Jiang2020Hub} The specific alignment direction of the stripes is so sensitive to $t'$ where even a small anisotropy in $t'$ can result in the subtle observed tilting. This highlights how the phases of stripes and superconductivity are sensitively intertwined at the level of model calculations and accounts for the appearance of these phases in a real material. Of course, the phenomenology of density wave ordering is different in LSCO compared to YBCO. This calls for a better understanding of how small changes in the Hamiltonian parameters, perhaps arising from subtle structural differences, can account for variations in the competing states across the families of cuprate materials.

\section{\label{sec:methods}Methods}
\subsection{\label{sec:methods_sample}Sample details}
The single crystal of LSCO with a nominal doping of $x=0.12$ was grown by the traveling-solvent-floating-zone method. As-grown crystal was annealed in oxygen gas flow at 900$^{\circ}$C for 72 h. At room temperature, the lattice constants are $a = b = 3.780$~\AA{} and $c = 13.218$~\AA{} determined by powder x-ray diffraction on a crushed piece of the single crystal. The structural transition temperature from the high-temperature tetragonal phase to the low-temperature orthorhombic phase is $T_s=242$~K determined by extinction of the nuclear Bragg peak (2,0,0) in neutron diffraction measurements upon warming. Both the lattice constants and the structural transition temperature $T_s$ agree pretty well with the previous reported results from powder and single crystal samples \cite{Radaelli1994,Yamada1998,Kimura2000}, reassuring the Sr content. To avoid the large mosaic that could be introduced from coalignment of multiple samples, only one large single crystal with a mass of $12.8$~g was used in our neutron scattering experiments. The size of the crystal is $8$~mm $\phi\times40$~mm.
\subsection{\label{sec:methods_neutron}Neutron measurements}
The neutron experiments were carried out at the thermal-neutron beamline HB-1A at the High Flux Isotope Reactor (HFIR) at Oak Ridge National Laboratory and the cold-neutron beamline at the Spin Polarized Inelastic Neutron Spectrometer (SPINS) at the Center for Neutron Research (NCNR) at the National Institute of Standards and Technology (NIST). For HB-1A, the incident energy was fixed at $14.65$~meV with horizontal collimations of $40'$-$40'$-sample-$40'$-$80'$. For SPINS, the final energy was fixed at $5$~meV, and most of measurements were done with horizontal collimations of guide-open-sample-$80'$-open, except for the mesh-scan near nuclear Bragg peak (1,1,0) (see Supplementary Fig.~1), which was with tighter horizontal collimations of guide-$20'$-sample-$20'$-open. The sample was mounted on an aluminum holder and aligned in the $(HK0)$ plane. Base temperatures of $T = 5$~K (HB-1A) and $T = 2.8$~K (SPINS) were achieved using closed cycle refrigerators. The sample was cooled slowly with a rate of $\sim2$~K per minute from room temperature.

Both the elastic scattering (Fig.~\ref{fig:1}) and temperature dependence of the inelastic scattering near (0.5,0.5,0) (Fig.~\ref{fig:4}) were measured at HB-1A. The elastic scattering near (1,1,0) (Supplementary Fig.~1) and (0.5,1.5,0) (Fig.~\ref{fig:2}) and energy dependence of the inelastic scattering near (0.5,0.5,0) (Fig.~\ref{fig:3}) were measured at SPINS. We note that the sample was rotated by 90$^{\circ}$ in-plane between the HB-1A and SPINS experiments, so the same structural domains between the two experiments were related by this 90$^{\circ}$ rotation. For example, the green domain measured at SPINS corresponds to the red one at HB-1A. This has been taken into account in the analysis of HB-1A data when domain population information is needed (including Supplementary Fig.~2 and the moment size calculation).

\subsection{\label{sec:methods_dmrg}Numerical calculations}%
We employed the \gls*{DMRG} \cite{White1992} method to study the ground state properties of the $t$-$t'$ Hubbard model as defined in Eq.(\ref{Eq:Ham}). We considered the square lattice with open boundary conditions in both directions specified by the basis vectors $\hat{x}=(1,0)$ and $\hat{y}=(0,1)$. The total number of sites is $N=16\times 8$, where $L_x=16$ and $L_y=8$ are the number of sites in $\hat{x}$ and $\hat{y}$ directions, respectively. The doped hole concentration is defined as $\delta=(N-N_e)/N$ with $N_e$ the number of electrons. Additional calculations were performed with smaller sample sizes ($N=12\times 8$ and $N=8\times 8$) at the same hole concentration. We performed around 60 sweeps in the current \gls*{DMRG} simulation and kept up to $m=25000$ number of states with a typical truncation error $\epsilon\approx 2\times 10^{-5}$.

\section*{Data availability}
All data needed to evaluate the findings in this paper are present in the paper and/or the Supplementary Information, and are available from the corresponding authors upon reasonable request.

\section*{Code availability}
The codes implementing the calculations of this study are available from the corresponding authors upon reasonable request.

\section*{References}
\bibliographystyle{naturemag}
\bibliography{ref}
\clearpage

\section*{Acknowledgments}
This work was supported by the U.S. Department of Energy, Office of Science, Basic Energy Sciences, Materials Sciences and Engineering Division, under contract DE-AC02-76SF00515. Research conducted at ORNL's High Flux Isotope Reactor was sponsored by the Scientific User Facilities Division, Office of Basic Energy Sciences, US Department of Energy. We acknowledge the support of the National Institute of Standards and Technology, U.S. Department of Commerce, in providing the neutron research facilities used in this work. M.F. was supported by Grant-in-Aid for Scientific Research (A) (Grant Nos. 16H02125 and 21H04448).

\section*{Author contributions}
The project was conceived by Y.S.L. W.H., J.W., G.X., and W.T.\ performed the neutron scattering measurements, and W.H.\ analyzed the data. T.T., Y.I., and M.F.\ synthesized the sample. H.C.J.\ performed the numerical calculations. The paper was written by W.H., H.C.J., and Y.S.L.\ with input from all co-authors.

\section*{Competing interests}
The authors declare no competing interests.

\clearpage

\end{document}

\title{Supplementary Information for: Tilted Stripes Origin in {La$_{1.88}$Sr$_{0.12}$CuO$_{4}$} Revealed by Anisotropic Next-Nearest Neighbor Hopping}

\author{Wei He}
 \email[]{Current address: Department of Condensed Matter Physics and Materials Science, Brookhaven National Laboratory, Upton, New York 11973, USA; whe1@bnl.gov}
 \affiliation{Stanford Institute for Materials and Energy Sciences, SLAC National Accelerator Laboratory, Menlo Park, CA 94025, USA}
 \affiliation{Department of Materials Science and Engineering, Stanford University, Stanford, CA 94305, USA}
\author{Jiajia Wen}
 \affiliation{Stanford Institute for Materials and Energy Sciences, SLAC National Accelerator Laboratory, Menlo Park, CA 94025, USA}
\author{Hong-Chen Jiang}
 \affiliation{Stanford Institute for Materials and Energy Sciences, SLAC National Accelerator Laboratory, Menlo Park, CA 94025, USA}
\author{Guangyong Xu}
 \affiliation{NIST Center for Neutron Research, National Institute of Standards and Technology, Gaithersburg, MD 20899-6102, USA}
\author{Wei Tian}
 \affiliation{Neutron Scattering Division, Oak Ridge National Laboratory, Oak Ridge, Tennessee 37831, USA}
\author{Takanori Taniguchi}
 \affiliation{Institute for Materials Research, Tohoku University, Sendai, 980-8577, Japan}
\author{Yoichi Ikeda}
 \affiliation{Institute for Materials Research, Tohoku University, Sendai, 980-8577, Japan}
\author{Masaki Fujita}
 \affiliation{Institute for Materials Research, Tohoku University, Sendai, 980-8577, Japan}
\author{Young S. Lee}
 \email[]{youngsl@stanford.edu}
 \affiliation{Stanford Institute for Materials and Energy Sciences, SLAC National Accelerator Laboratory, Menlo Park, CA 94025, USA}
 \affiliation{Department of Applied Physics, Stanford University, Stanford, CA 94305, USA}

\date{\today}% It is always \today, today,
             %  but any date may be explicitly specified

\maketitle

\tableofcontents

\clearpage

\section{Structural Twinning and Domain Populations}
\begin{figure}[htbp]
 \includegraphics[scale=1]{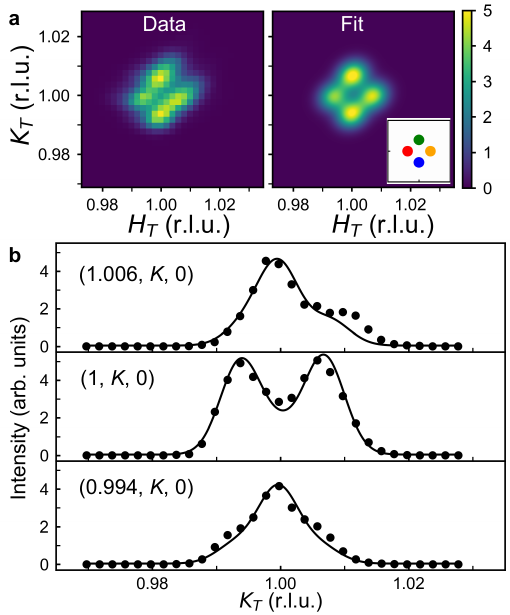}%
 \caption{\textbf{Structural twinning evidenced by the nuclear Bragg peak (1,1,0)}. \textbf{a}, Mesh scan near the nuclear Bragg peak (1,1,0) measured at $T=2.8$~K at SPINS (left) and its 2D fit (right) as described in the text. The expected positions for these peaks are drawn in the inset with the same color code used in the main figures. \textbf{b}, Representative cuts along $K$ direction to show the good match between the data and the fit. Error bars correspond to $\pm \sigma$, where $\sigma$ is the standard deviation.}
 \label{fig:S1}
\end{figure}

Structural transitions are often accompanied by twinning phenomena, which can restore the broken high-temperature symmetry at macroscopic scale. As depicted in Fig.~1\textbf{b}, four possible twin domains—forming two pairs, each with $\{ 100 \}$ planes in the high-temperature tetragonal (HTT) phase as the mirror boundary—can coexist in the low-temperature orthorhombic (LTO) phase \cite{Braden1992}. Because of this geometric constraint, each domain has its twin counterpart that has nearly overlapped orthorhombic axes but offsets by a small angle the same as the orthorhombicity $\theta_{ortho}$. Although this twinning effect complicates the data analysis due to the superposition of four sets of peaks in reciprocal space, no information is lost as long as we have enough resolution to resolve equivalent peaks from different domains. For example, by measuring the nuclear Bragg peak (1,1,0), we are able to determine the domain populations in our sample. As shown in Supplementary Fig.~\ref{fig:S1}\textbf{a}, each domain contributes to one individual peak and these four peaks are well resolved with the tight collimation setting. After fitting with four delta functions (in both momentum and energy) convolved with the instrumental resolution, we obtain peak intensities, which are proportional to the domain populations (Supplementary Table~\ref{table:S1}). Besides, the sample mosaic is determined to be $13.3(2)'$ (\gls*{FWHM}) from the fit, and the orthorhombicity angle $\theta_{ortho}$ calculated from the fitted peak positions is $0.38(2)^{\circ}$.

\begin{table}[htpb]
\caption{\textbf{Structrual twinning information}. Fitted peak position and intensity of the nuclear Bragg peak (1,1,0) for each structural twin domain and the corresponding domain population.}
\renewcommand\arraystretch{1}
\begin{ruledtabular}
\begin{tabular}{ccccc}
\multirowcell{2}{\textbf{Domain}} & \multirowcell{2}{$\bm{H}$\\ \textbf{(r.l.u.)}} & \multirowcell{2}{$\bm{K}$\\ \textbf{(r.l.u.)}}  & \multirowcell{2}{\textbf{Intensity}\\ \textbf{(arb. units)}}  & \multirowcell{2}{\textbf{Population} \\ \textbf{($\%$)}} \\
 & & & & \\
\hline
\textbf{Red} & 0.9938(1) & 0.9996(1) & 57.3(9) & 21.1(3) \\
\textbf{Blue} & 0.9999(1) & 0.9938(1) & 70.8(9) & 26.1(3) \\
\textbf{Green} & 1.0003(1) & 1.0069(1) & 76.3(9) & 28.1(3) \\
\textbf{Orange} & 1.0072(1) & 0.9999(1) & 67.0(9) & 24.7(3) \\
\end{tabular}
\end{ruledtabular}
\label{table:S1}
\end{table}

\clearpage

\section{Elastic Scattering of the Static Spin Stripes}
\subsection{Index of the \gls*{SDW} peaks in orthorhombic notation}
Figure~1\textbf{d} shows the measurement of the \gls*{SDW} peaks around (0.5,0.5,0). The peak center in each scan deviates from 0.5, indicating the existence of the \textit{Y}-shift. However, this observation is still quite different from the predicted pattern based on the presence of four structural domains shown in Fig.~1\textbf{c}. The four twin domains can be grouped into two pairs (i.e., the blue-red pair and the orange-green pair drawn in Fig.~1\textbf{b}, each containing a lattice rotating from its twin counterpart by $\theta_{ortho}$ of merely $0.38^{\circ}$, beyond the instrument resolution we used here. Even though considering such resolution effect, there should still be two resolvable peaks in each scan due to the \textit{Y}-shift, with the averaged center at $H(K)=0.5$. Careful inspection reveals that the four observed peaks actually share a common center with $H$ and $K$ coordinates slightly larger than $0.5$. This center corresponds to the (100) position (instead of (010)) in orthorhombic notation, since $a_{\rm o}<b_{\rm o}$. Therefore, these peaks must originate from the orange and green domains, which have their orthorhombic $\mathbf{a}_{\rm o}$ axis nearly aligned with this center. Then using the determined orthorhombic lattices, we can index the observed peaks as ($1\pm\delta_h$,$\pm\delta_k$,0) in orthorhombic notation and find $\delta_h=0.1111(6)$ and $\delta_k=0.1234(6)$. This anisotropic incommensurabilities with $\delta_h<\delta_k$ indicate that the \textit{Y}-shift angle is opposite to the orthorhombic distortion, as drawn in Fig.~1\textbf{a}.

\subsection{In-plane correlation length of the \gls*{SDW} order}

\begin{figure}[htbp]
 \includegraphics[scale=1]{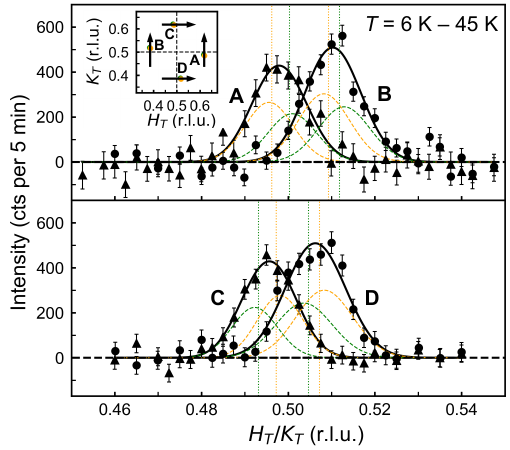}%
 \caption{\textbf{Elastic neutron scattering of the \gls*{SDW} peaks measured at $T=6$~K at HB-1A, where the $T=45$~K data was subtracted as background}. The solid lines are fits to the data after taking into account the instrumental resolution, as described in the text. The dashed lines show the contributions from individual domains with the centers at the vertical dotted lines. The trajectory of each scan is denoted in the inset. Error bars correspond to $\pm \sigma$, where $\sigma$ is the standard deviation.}
 \label{fig:S2}
\end{figure}

In order to extract the in-plane correlation length of the \gls*{SDW} order, we use the data of the \gls*{SDW} peaks around (0.5,0.5,0) measured at HB-1A (Fig.~1), which give us the best statistics, and perform fits with consideration of instrumental resolution.  As shown in Supplementary Fig.~\ref{fig:S2}, the solid lines are the results of fits to \gls*{2D} Gaussian peaks (with a delta function in energy) convolved with the instrumental resolution. This model is based on the fact that the out-of-plane correlation length is extremely short. For each scan, two peaks are used to model the contributions from the two observable structural twin domains, whose crystal axes offset from each other by $\theta_{ortho}$. The widths of the two peaks are constrained to be equal. Their relative positions are fixed to the projection of the orthorhombic splitting on the scan direction, and the relative intensities are fixed based on the domain populations. Supplementary Table~\ref{table:S2} lists the extracted correlation lengths (calculated as 2/FWHM), with a mean value of $123(7)$~\AA{}.

\begin{table}[htbp]
\caption{\textbf{In-plane correlation lengths of the \gls*{SDW} order}. These values are extracted from the fits in supplementary Fig.~\ref{fig:S2}.}
\renewcommand\arraystretch{1}
\begin{ruledtabular}
\begin{tabular}{ccc}
\multirowcell{2}{\textbf{Scan position}} & \multirowcell{2}{\textbf{Correlation}\\ \textbf{length (\AA{})}} & \multirowcell{2}{\textbf{Mean correlation}\\ \textbf{length (\AA{})}} \\
 & & \\
\hline
\textbf{A} & 113(16) & \multirowcell{4}{123(7)}  \\
\textbf{B} & 123(15) &  \\
\textbf{C} & 138(14) &  \\
\textbf{D} & 118(13) &  \\
\end{tabular}
\end{ruledtabular}
\label{table:S2}
\end{table}

\subsection{Fitting method for the \gls*{SDW} peaks around (0.5,1.5,0)}
Figure~2 displays the fits to the \gls*{SDW} peaks around (0.5,1.5,0) measured at SPINS based on the two proposed models, i.e., the 3D stacking model and Spin $\parallel \mathbf{b}_{\rm o}$ model. Similar to the fits in Supplementary Fig.~\ref{fig:S2}, we use a \gls*{2D} Gaussian peak (with a delta function in energy) to model the cross section from each structural domain and convolve it with the instrumental resolution. The final curve is a superposition of four peaks from the four structural domains. 
We use the information learned from the \gls*{SDW} peaks around (0.5,0.5,0) to constrain the fits here. For each Gaussian peak, the peak position is fixed to the value predicted by the previously determined \textit{Y}-shift angle $\theta_{Y}$ and orthorhombicity angle $\theta_{ortho}$, and the peak width is also fixed to give the same in-plane correlation length. 

For the 3D stacking model, we assume the correlation length along $L$ direction is longer than $14.3$~\AA{}.  In this case, the peaks from the orange and green domains will be negligible and we set their intensities to zero in the fit. The ratio of the intensities between the other two domains (i.e., red and blue) is also fixed to the value from the structural domain populations, so the only fitting parameter here is the overall intensity. 

For the Spin $\parallel \mathbf{b}_{\rm o}$ model, we further utilize the information of the peak intensities near (0.5,0.5,0) to constrain the fit. This is done by co-fitting with the \gls*{SDW} peak measured near (0.5,0.5,0) in the same experiment, assuming that the differences of \gls*{SDW} peak cross sections between the two Brillouin zones come only from the magnetic form factor (which can be calculated from the tabulated data \cite{magneticFormFactor}), correlation length along $L$ direction and the spin polarization factor (which can be determined by the relative angle between the wave vector $\mathbf{Q}$ and the spin direction). Since, the peak measured near (0.5,0.5,0) are only from the green and orange domains, this co-fit can only constrain the peak intensities from these two domains near (0.5,1.5,0) with a fudge factor introduced by a short but non-zero correlation length along $L$. For the other two domains (i.e., red and blue), this interlayer correlation length has no effect on their intensities. We fix the ratio of the intensities between the green and orange domains to be the same as the structural domain populations, since we do not have enough resolution to resolve these two peaks. And do the same for the other two domains. But we let the ratio of the intensities between these two pairs to be freely refined, and find that this ratio is $2.9(6)$. This indicates that the volume fractions of stripe order phases are different in the four structural twin domains. As shown in the comparison plots in Supplementary Fig.~\ref{fig:population_imbalance}, it is indeed necessary to introduce such non-uniform stripe volume fraction in the fits of \gls*{SDW} peaks around (0.5,1.5,0). Otherwise, the intensities from the red and blue domains cannot be well described.

\begin{figure}[htbp]
 \includegraphics[scale=0.9]{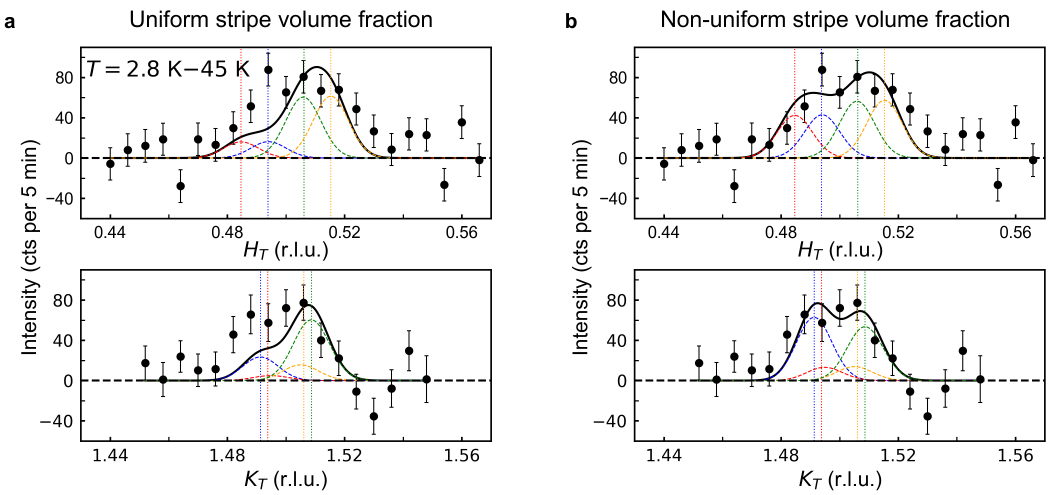}%
 \caption{\textbf{The effect of non-uniform stripe volume fraction on the fits for \gls*{SDW} peaks around (0.5,1.5,0)}. The data are the same as those in Fig.~2\textbf{e} or \textbf{f}, which show elastic neutron scattering of \gls*{SDW} peaks around (0.5,1.5,0) at $2.8$~K with the $45$~K data subtracted as a background. The fits in \textbf{a} assume uniform stripe volume fraction for spin stripes, i.e., the stripe populations are the same as the structural domain populations revealed by the nuclear Bragg peaks. The fits in \textbf{b} are the same as those in Fig.~2\textbf{f}, which introduce a fitting parameter to account for the non-uniform stripe volume fraction in the two pairs of structural twin domains. The dashed curve shows the component from each peak with the center at the vertical dotted line. Error bars correspond to $\pm \sigma$, where $\sigma$ is the standard deviation.}
 \label{fig:population_imbalance}
\end{figure}

\subsection{The finite-size domain model for the calculation of interlayer correlation length}
Reference~\cite{Lee1999} employed a finite-size domain model to estimate the interlayer correlation length in oxygen-doped \ce{La_$2$CuO_$4+y$} (LCO). For a direct comparison, we use the same method in this paper to evaluate the interlayer correlation length. In this model, a Gaussian function is used to estimate the peak broadening due to a finite domain size \cite{x-ray}. The domain size (correlation length) is related to the \gls*{FWHM} of the peak by $\xi = 4\sqrt{\pi \ln{2}}/\mathrm{FWHM}\approx5.9/\mathrm{FWHM}$.

\subsection{Comparison of stripe stacking arrangement between LSCO and LBCO}
\begin{figure}[htbp]
 \includegraphics[scale=1]{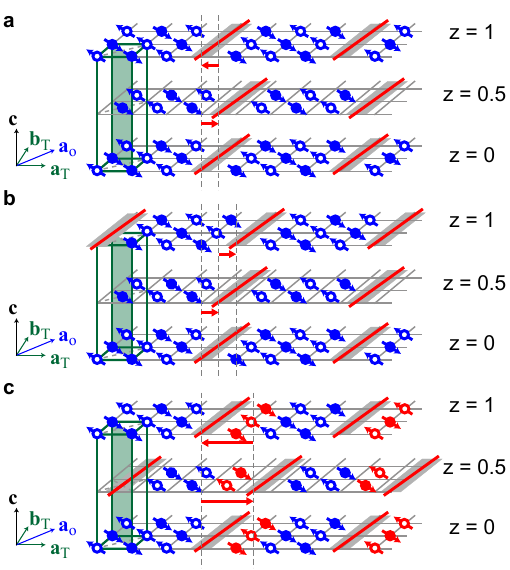}%
 \caption{\textbf{Possible stripe stacking arrangements in LSCO}. The domain walls on neighboring layers are shifted by 0.5 of a lattice spacing in \textbf{a} and \textbf{b}, and shifted by 1.5 lattice spacings in \textbf{c}. The solid green lines denote the structural unit cell. The grey regions are the charge stripes and the red lines show the tilted effective stripe direction. Solid and open circles denote spins with opposite directions, and arrows on top of the circles further specify spin directions. The horizontal red arrows indicate both the direction and magnitude of the shift of charge domain walls between neighboring layers. The red spins in \textbf{c} are in the higher energy configuration, and therefore the stacking arrangement in \textbf{c} is less likely for LSCO.
 \label{fig:stacking}
}
\end{figure}

Compared to the materials with low-temperature tetragonal (LTT) structure, such as LBCO, the correlation length of the charge stripe order peaks in LSCO has been found to be shorter \cite{Croft2014}. In contrast to the relatively sharp peaks at half-integer $L$ positions in LBCO \cite{Kim2008}, the charge order peaks in LSCO do not show similarly strong correlations along $L$ \cite{Croft2014}. Here, we find that the spin stripe order in LSCO has a fairly short correlation length of $\sim7$~\AA~along $L$, comparable with its charge correlation length. These observations can be understood in terms of the different stripe stacking arrangements in these two types of materials.

For La-based cuprates, there are two \ce{CuO2} layers in each structural unit cell with a body-centered arrangement. For materials with a crystal structure similar to LBCO, the LTT structure can pin the direction of the stripes resulting in orthogonal stripes on the neighboring layers. Then the Coulomb repulsion between charge stripe across second neighbor planes can give a stacking arrangement with a doubled unit cell along the $c$-axis, resulting in charge order peaks at half-integer $L$, as seen in (LaNd)$_{2-x}$Sr$_x$CuO$_4$.\cite{Tranquada2013}

However, the stacking arrangement in LSCO is different due to the presence of the orthorhombic distortion. The distortion lifts the frustration of the nearest neighbor magnetic coupling between layers. Interlayer nearest neighbor spins are favored to align antiparallel, which is the case in the parent compound \ce{La$_2$CuO$_4$} \cite{Vaknin1987}. Our LSCO sample has a similar orthorhombic distortion and spin ordering direction as the parent compound. For the stacking of the stripes, there is a compromise between Coulomb repulsion and the exchange interaction between spins. Coulomb repulsion favors charge domain walls with large spacing across the layers, while the magnetic coupling favors antiferromagnetically aligned spins between nearest neighbor interlayer sites. If the interlayer magnetic coupling term plays the more important role, then the charge domain walls would align nearly on top of each other as shown in Supplementary Fig.~\ref{fig:stacking}\textbf{a} and \textbf{b}. However, if there is a significant shifting of the charge stripe locations between neighboring layers as drawn in Supplementary Fig.~\ref{fig:stacking}\textbf{c}, there will be unsatisfied spins shown in red. For the same reason, orthogonal interlayer stripes are also not favorable in LSCO. The $L$-dependence of the spin stripe order can help distinguish these cases. Indeed, prior results from oxygen-doped LCO provide compelling evidence \cite{Lee1999} to support the stacking arrangement shown in Supplementary Fig.~\ref{fig:stacking}\textbf{a} and \textbf{b} instead of \textbf{c}. Our results are consistent with the same stacking arrangement as in oxygen-doped LCO, albeit with a shorter correlation length along the $c$-axis ($\sim7$~\AA~ versus $\sim13$~\AA~in oxygen-doped LCO).

The short $c$-axis correlation length may arise from stacking faults. Though the charge domain walls are nearly on top of each other between adjacent layers, due to the body-centered structure, there are two equivalent shift directions for the adjacent layer (left or right by half of the lattice spacing). For three layers, there are four permutations of the stripe stackings, and supplementary Fig.~\ref{fig:stacking}\textbf{a} and \textbf{b} only show two of them. If the Coulomb interaction between stripes on further neighbor layers is not strong, such stacking faults may suppress stripe correlations along the $c$ axis and give broad spin and charge ordering peaks along $L$. We note that the above discussion is based on site-centered stripes, but can also apply to the bond-centered case.

\clearpage

\section{Inelastic Scattering of the Fluctuating Spin Stripes}
\subsection{Ruling out the possibility that the inelastic signal is only from the stripe ordered region}

\begin{figure}[htbp]
 \includegraphics[scale=1]{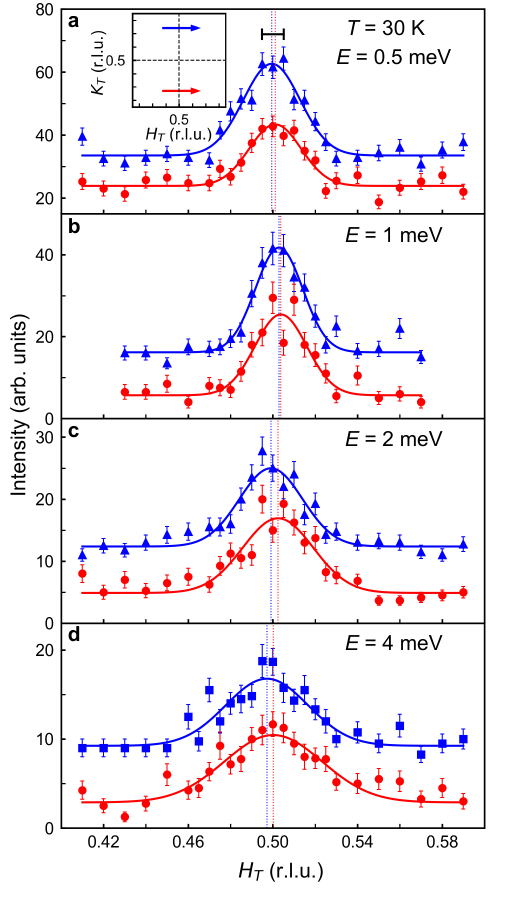}%
 \caption{\textbf{Inelastic neutron scattering of the spin stripes at $T=30$~K with various energy transfers}. The solid lines are Gaussian fits with constant background and the vertical dotted lines are the fitted peak centers. The trajectory of each scan is denoted in the inset of \textbf{a}. The horizontal bar represents the elastic peak shift. The blue data are shifted vertically for clarity. Error bars correspond to $\pm \sigma$, where $\sigma$ is the standard deviation.
}
\label{fig:S3}
\end{figure}

Figure~3\textbf{a} and \textbf{b} shows the dynamic spin stripes measured at $E = 0.5$~meV and $E = 2$~meV at SPINS. Additional data at higher energy transfers are shown in Supplementary Fig.~\ref{fig:S3}. To test the possibility that the inelastic scattering is from the same minor region where the elastic signal is from, we calculate the weighted peak positions for isotropic and transverse excitations, respectively, assuming that the inelastic intensities from different structural domains are proportional to the magnetically ordered volumes revealed from the fits in Fig.~2\textbf{f}. Specifically, the ordered stripe volume fraction in the red and blue domains are $2.9$ times larger than those in the green and orange domains. As shown in Supplementary Fig.~\ref{fig:S4}, this non-uniform stripe volume fraction results in large negative peak shift, clearly deviated from the observation. Thus, we can rule out this possibility. In fact, the measured values are better described by isotropic spin fluctuations with \textit{Y}-shift from the whole sample indicated by the solid black line in Supplementary Fig.~\ref{fig:S4}.

\begin{figure}[htbp]
 \includegraphics[scale=1]{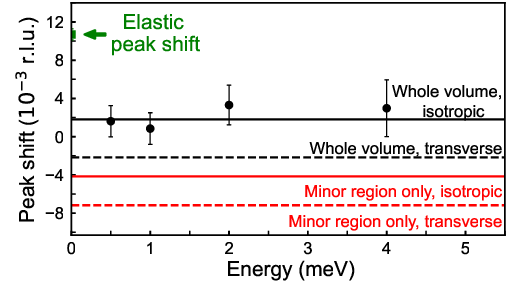}%
 \caption{\textbf{Energy dependence of the peak shift}. This shift is measured between the fitted centers of a pair of peaks as shown in Fig.~3\textbf{a}--\textbf{b} and Supplementary Fig.~\ref{fig:S3} with the expected values of the peak shift based on different assumptions. Solid and dashed red lines are the expected peak shift for isotropic and transverse excitations with \textit{Y}-shift, respectively, assuming the inelastic signal is only from the minor region with static spin stripe order; while the solid and dashed black lines are reproduced from the Fig.~3\textbf{c} to show the expected peak shift for isotropic and transverse excitations with \textit{Y}-shift, respectively, assuming the inelastic signal is from the whole sample. Error bars correspond to $\pm \sigma$, where $\sigma$ is the standard deviation.
}
\label{fig:S4}
\end{figure}

\subsection{Dynamical spin correlation length}

We perform the following analysis on the inelastic scans at $30$~K with an energy transfer of $E = 0.5$~meV measured at SPINS to extract the in-plane dynamical spin correlation length. As shown in Supplementary Fig.~\ref{fig:S5}, the solid lines are the results of fits to \gls*{2D} Gaussian peaks (constant in energy) convolved with the instrumental resolution. For each scan, four peaks are used to model the contributions from the four structural twin domains. The widths of the four peaks are constrained to be equal and the peak positions are fixed to the expected values assuming the same \textit{Y}-shift angle in the inelastic scattering. Their relatively intensities are also fixed based on three assumptions. First, spin excitations exist in the whole volume of the sample. Second, the excitations are isotropic. This should only apply to the non-magnetically ordered regions, but since the magnetic volume fraction is very small ($\sim4\%$), we neglect such differences and treat them the same. Last, interlayer correlation length is the same as the elastic scattering. This causes a drop of intensity by $\sim20\%$ in the red and blue domains, but barely affects the fitted peak width. The dynamical correlation length is calculated as 2/FWHM. The scan in panel \textbf{a} gives better statistics, and the extracted dynamical correlation length is $62(10)$~\AA{}. The scan in panel \textbf{b} will give rather large error bars for the fitted peak widths. After we fix the widths to be the same as those in scan \textbf{a}, we indeed obtain satisfactory fitting result shown as the solid line, which can serve as a cross-check.

\begin{figure}[htbp]
 \includegraphics[scale=1]{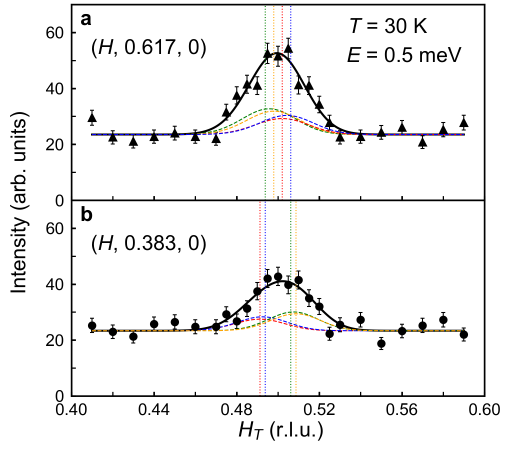}%
 \caption{\textbf{Inelastic neutron scattering of the spin stripes at $30$~K with an energy transfer of $E = 0.5$~meV measured at SPINS}. Scan \textbf{a} is at $K = 0.617$ and \textbf{b} at $K = 0.383$. The solid lines are fits to the data after taking into account the instrumental resolution, as described in the text. The dashed lines show the contributions from individual domains with the centers at the vertical dotted lines. Error bars correspond to $\pm \sigma$, where $\sigma$ is the standard deviation.}
 \label{fig:S5}
\end{figure}

\subsection{Fitting method for the inelastic data in the comparison between isotropic- and transverse- excitation models}
The inset of Fig.~3\textbf{c} displays the fits to the inelastic scan at $K=0.383$ based on the isotropic- and transverse- excitation models. We choose this data set instead of the scan at $K=0.617$. This is because the expected peak positions of the four structural domains are more dispersed in the scan at $K=0.383$ (see the vertical lines in Supplementary Fig.~\ref{fig:S5}), which makes it easier to distinguish between the two models. The fitting method is similar to the way we extract the dynamical correlation length in Supplementary Fig.~\ref{fig:S5}. We again use four peaks to fit the data. Each peak is a Gaussian peak in \gls*{2D} (constant in energy) convolved with the instrumental resolution. The peak positions are fixed to the expected values and the width are also fixed to give the same dynamical correlation length ($62$~\AA{}) we obtained above. The relative intensities are fixed assuming either isotropic or transverse excitations in the whole volume of the sample. It should be noted that the fitting result is not sensitive to the fixed value of peak width we choose in the fit, but mainly governed by the relative intensities reflecting the difference between the two models.

\subsection{Comparison of the expected \textit{Y}-shift between single-domain and multi-domain samples}
As shown in Supplementary Fig.~\ref{fig:simulated_Yshift}, the single-domain case has a larger slope than multi-domain cases and gives the same peak shift as the elastic one. Once there are multiple domains, the averaging effect reduces the peak shift. However, the difference between isotropic- and transverse- excitations always exists as long as the intrinsic \textit{Y}-shift angle is non-zero. And the consistently positive shifts in our inelastic measurements clearly favor the isotropic excitation case with non-zero intrinsic \textit{Y}-shift. We also perform fitting based on this scenario, and the fitted intrinsic \textit{Y}-shift is $3.1(9)^{\circ}$, which, within the errors, is the same as the tilt angle in elastic scattering.

\begin{figure}[htbp]
 \includegraphics[scale=1]{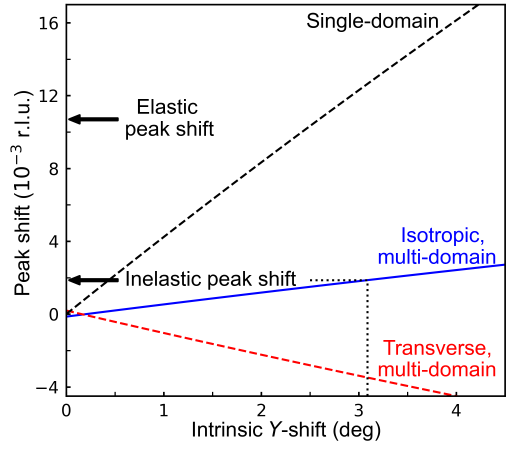}%
 \caption{\textbf{Comparison of the simulated \textit{Y}-shift between single-domain and multi-domain samples}. The x-axis is the intrinsic \textit{Y}-shift angle for fluctuating spin stripes. The y-axis is the simulated peak shift between a pair of inelastic peak positions as shown in the inset of Fig.~3\textbf{a}. The arrow labeled with “inelastic peak shift” is determined by the average shift value of the four energy transfers shown in Fig.~3\textbf{c}. The black dashed line is calculated for single-domain samples, while the blue (red) lines are for multi-domain samples with domain populations the same as in our sample and with isotropic (transverse) excitations.}
 \label{fig:simulated_Yshift}
\end{figure}

\subsection{Temperature dependence of the dynamic spin stripes}
\begin{figure}[htbp]
 \includegraphics[scale=1]{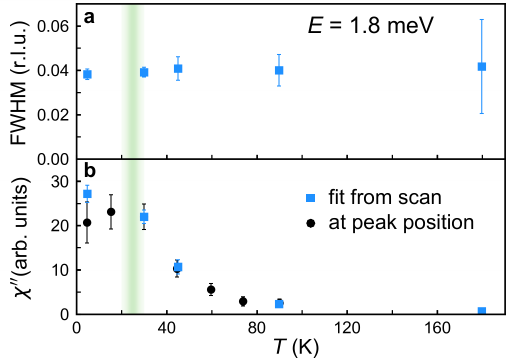}%
 \caption{\textbf{Temperature dependence of the dynamic spin stripes at an energy transfer of $E = 1.8$~meV measured at HB-1A}. \textbf{a}, \gls*{FWHM} extracted from the Gaussian fits of the temperature dependence measurements. \textbf{b}, The imaginary part of the susceptibility $\chi''$ extracted from Fig.~4\textbf{b}. The black circles are from one-point measurement at the peak position. The diffuse green vertical line indicates the \gls*{SDW} onset temperature varying from $\sim20$~K ($\mu$SR \cite{Savici2002}) to $\sim30$~K (neutron \cite{Kimura1999a}). Error bars correspond to $\pm \sigma$, where $\sigma$ is the standard deviation.}
 \label{fig:S6}
\end{figure}

Supplementary Fig.~\ref{fig:S6} shows the temperature dependence of the \gls*{FWHM} and imaginary part of the susceptibility $\chi''$ for the dynamic spin stripes measured with an energy transfer of $E = 1.8$~meV. The peak widths are found to be independent of temperature within the errors.

\clearpage

\section{Detailed Results of Numerical Simulations}

\subsection{Fitted \textit{Y}-shift of charge and spin ordering peaks}
\begin{figure}[htbp]
 \includegraphics[scale=0.85]{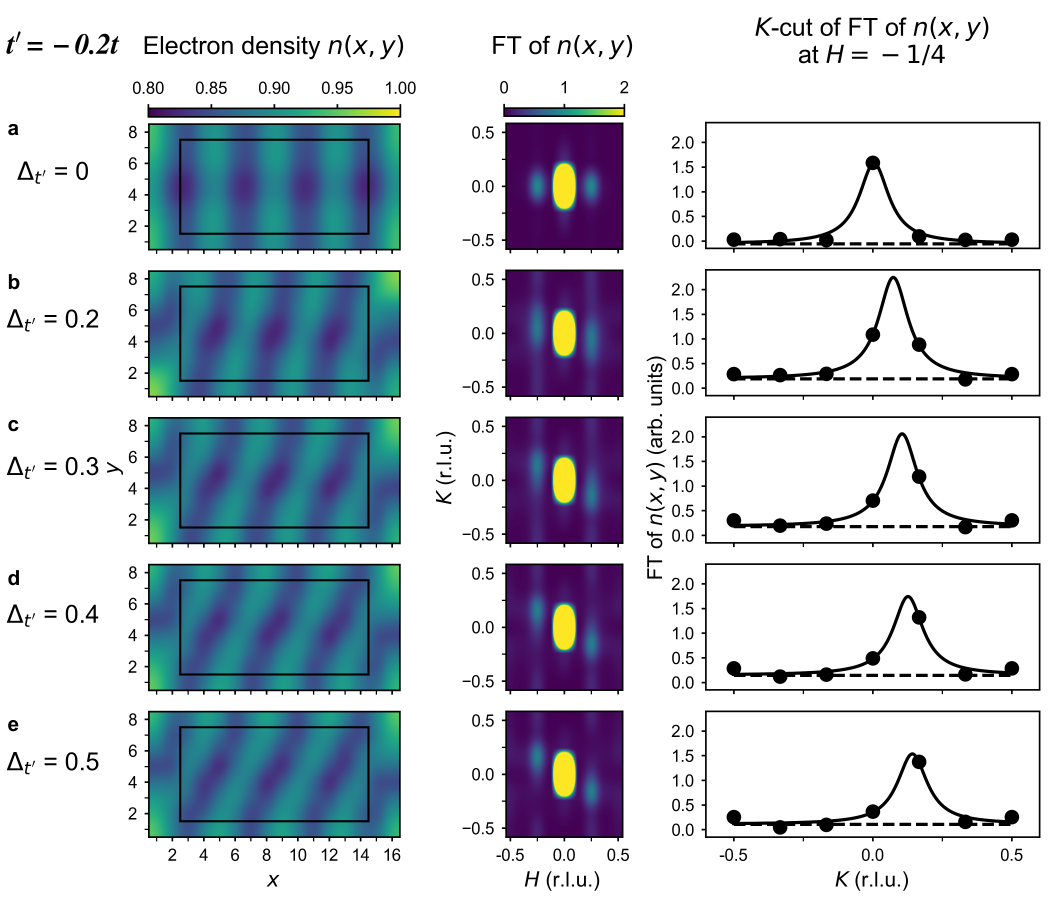}%
 \caption{\textbf{Charge stripes for $t'=-0.2t$ at various $\Delta_{t'}$ values calculated using the \gls*{DMRG} method}. This includes the electron density distributions $n(x,y)$ (left column), the corresponding Fourier transform (middle column) and $K$-cuts of the charge ordering peaks at $H=-0.25$ (right column). The rectangles in the left column depict regions used in the Fourier transform. The solid lines in the right column are Lorentzian fits with a constant background (dashed lines). The maps in the left and middle columns are produced using multiquadric interpolation method \cite{matplotlib}.}
 \label{fig:charge1}
\end{figure}

\begin{figure}[htbp]
 \includegraphics[scale=1]{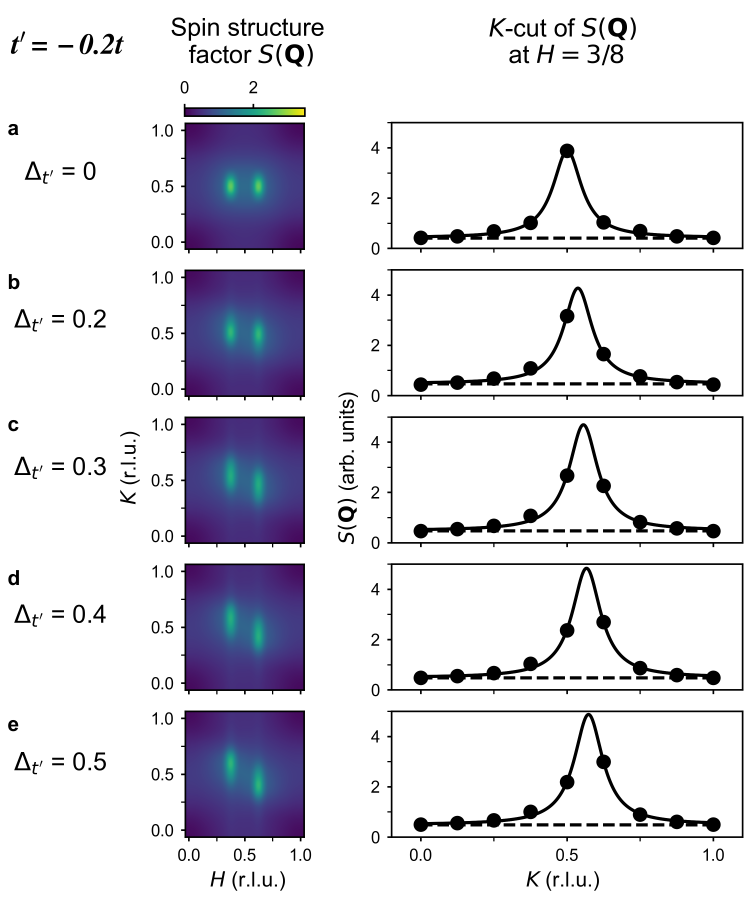}%
 \caption{\textbf{Spin stripes for $t'=-0.2t$ at various $\Delta_{t'}$ values calculated using the \gls*{DMRG} method}. This includes the static spin structure factors $S(\mathbf{Q})$ (left column) and $K$-cuts of the spin ordering peaks in $S(\mathbf{Q}$) at $H=0.375$ (right column). The solid lines in the right column are Lorentzian fits with a constant background (dashed lines). The maps in the left column are produced using multiquadric interpolation method \cite{matplotlib}.}
 \label{fig:spin1}
\end{figure}

Supplementary Figs.~\ref{fig:charge1} and \ref{fig:spin1} display the electron density distributions $n(x,y)$ and static spin structure factors $S(\mathbf{Q})$ for $t'=-0.2t$ at various $\Delta_{t'}$ obtained in \gls*{DMRG} \cite{White1992} calculations. From the Fourier transform of the electron density, we can also obtain the charge ordering peaks in reciprocal space (shown in the middle column of Supplementary Fig.~\ref{fig:charge1}). In order to determine the \textit{Y}-shift for both charge and spin ordering peaks, we take cuts along $K$ direction and fit the peaks with a Lorentzian line shape. Due to the finite sample size along $y$ direction, we have limited data points in these cuts. To get more accurate determination of the peak positions, the peak width is fixed to the best overall value based on data with all different $\Delta_{t'}$. The \textit{Y}-shift, i.e., the peak shift along $K$ direction relative to the case of $\Delta_{t'} = 0$, is summarized in Supplementary Fig.~\ref{fig:fit1}. We can see that the charge ordering peak is approximately commensurate with the spin ordering peak, i.e., $\Delta q_c \approx 2 \Delta q_s$. The small deviation can be attributed to the finite-size and boundary effects. Since charge stripes are more susceptible to such effects due to the Friedel oscillations from the boundary \cite{White2002}, we only use the fitted spin ordering peaks in the later estimation of the $\Delta_{t'} $ value in LSCO in the main text. 

Similarly, Supplementary Figs.~\ref{fig:charge2} and \ref{fig:spin2} show the electron density distributions $n(x,y)$ and static spin structure factors $S(\mathbf{Q})$ for $\Delta_{t'}=0.2$ at various $t'$ obtained in \gls*{DMRG} calculations, and the corresponding \textit{Y}-shift for both charge and spin ordering peaks are summarized in Supplementary Fig.~\ref{fig:fit2}

\subsection{\textit{Y}-shift and kinetic energy difference along the two diagonal bond directions}
Supplementary Fig.~\ref{fig:YshiftvsEnergydiff} shows the peak shift between a pair of spin ordering peaks in $S(\mathbf{Q})$ maps as a function of the kinetic energy difference between the $(1-\Delta_{t'})$-bonds and the $(1+\Delta_{t'})$-bonds. Interestingly, we note that this peak shift is proportional to the kinetic energy difference between the two diagonal bond directions, demonstrating the intimate relationship between the kinetic energy term and the \textit{Y}-shift phenomenon.

\subsection{More calculations with different system sizes}
We performed more calculations with smaller system sizes ($N=12\times 8$ and $N=8\times 8$) to show that the boundary effects and finite-size effects have negligible impacts on our above conclusions which are based on the system size of $N=16\times 8$. 

Supplementary Fig.~\ref{fig:charge_comparision} shows a side-by-side comparison of the electron density distributions $n(x,y)$ in the three systems with different sizes, including the Fourier transform of the electron density and the corresponding line cuts through the charge ordering peaks. Similarly, Supplementary Fig.~\ref{fig:spin_comparision} shows a comparison of the
static spin structure factors $S(\mathbf{Q})$ in these three systems with different sizes, together with the corresponding line cuts through the spin ordering peaks. Supplementary Figs.~\ref{fig:fit_comparision1} and ~\ref{fig:fit_comparision2} summarize the fitted \textit{Y}-shift values for both charge and spin ordering peaks with different system sizes. 

From the Supplementary Fig.~\ref{fig:fit_comparision1}\textbf{d}--\textbf{f}, we find that the mutual commensurate relationship between charge and spin stripes, manifest in the ratio $\Delta q_c / \Delta q_s \approx 2$, is satisfied starting from lower $\Delta_{t'}$ for larger system size, consistent with the trend that larger systems suffer from weaker finite-size and boundary effects. For $N=16\times 8$, this commensurate relationship maintains down to at least $\Delta_{t'}=0.2$. As shown in Supplementary Fig.~\ref{fig:fit_comparision2}, with the increase of the sample size, the fitted peak positions from the $N=12\times 8$ cluster match that in the $N=16\times 8$ cluster very well, especially for the spin ordering peaks. This suggests that the boundary effects and finite-size effects are negligible in extracting the \textit{Y}-shift once the system size is larger than $N=12\times 8$.

We noticed that the spin ordering peaks in the $N=12\times 8$ cluster are slightly broader than those in the other two system sizes, which could be due to the even-odd effect, as the $N=12\times 8$ cluster contains 1.5 \gls*{SDW} until cells, compared with the 2 \gls*{SDW} unit cells in the $N=16\times 8$ cluster and 1 \gls*{SDW} unit cell in the $N=8\times 8$ cluster. However, the peak positions are more robust and not obviously affected by this effect.

\begin{figure}[htbp]
 \includegraphics[scale=1]{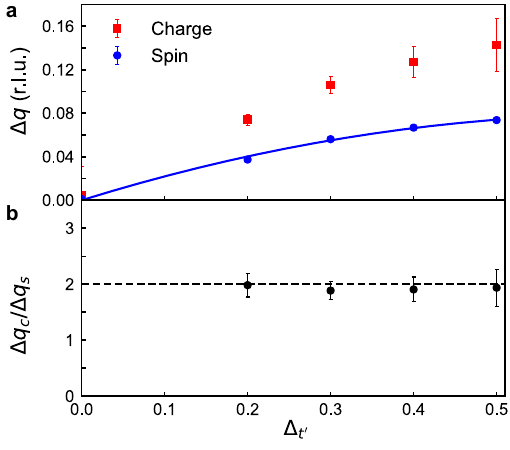}%
 \caption{\textbf{\textit{Y}-shift for both charge and spin ordering peaks for $t'=-0.2t$ from \gls*{DMRG} calculations and their ratios at various $\Delta_{t'}$ values}. $\Delta q_c$ ($\Delta q_s$) is the fitted charge (spin) peak positions along $K$ direction shown in the right column of Supplementary Fig.~\ref{fig:charge1} (\ref{fig:spin1}) relative to the position if $\Delta_{t'}=0$, i.e., $\Delta q_c=|q_c-0|$ ($\Delta q_s=|q_s-0.5|$). The solid blue line in \textbf{a} is a quadratic fit to the data and is the same as the fitted line in Fig.~6. Error bars correspond to $\pm \sigma$, where $\sigma$ is the standard deviation.
 }
 \label{fig:fit1}
\end{figure}

\begin{figure}[htbp]
 \includegraphics[scale=0.85]{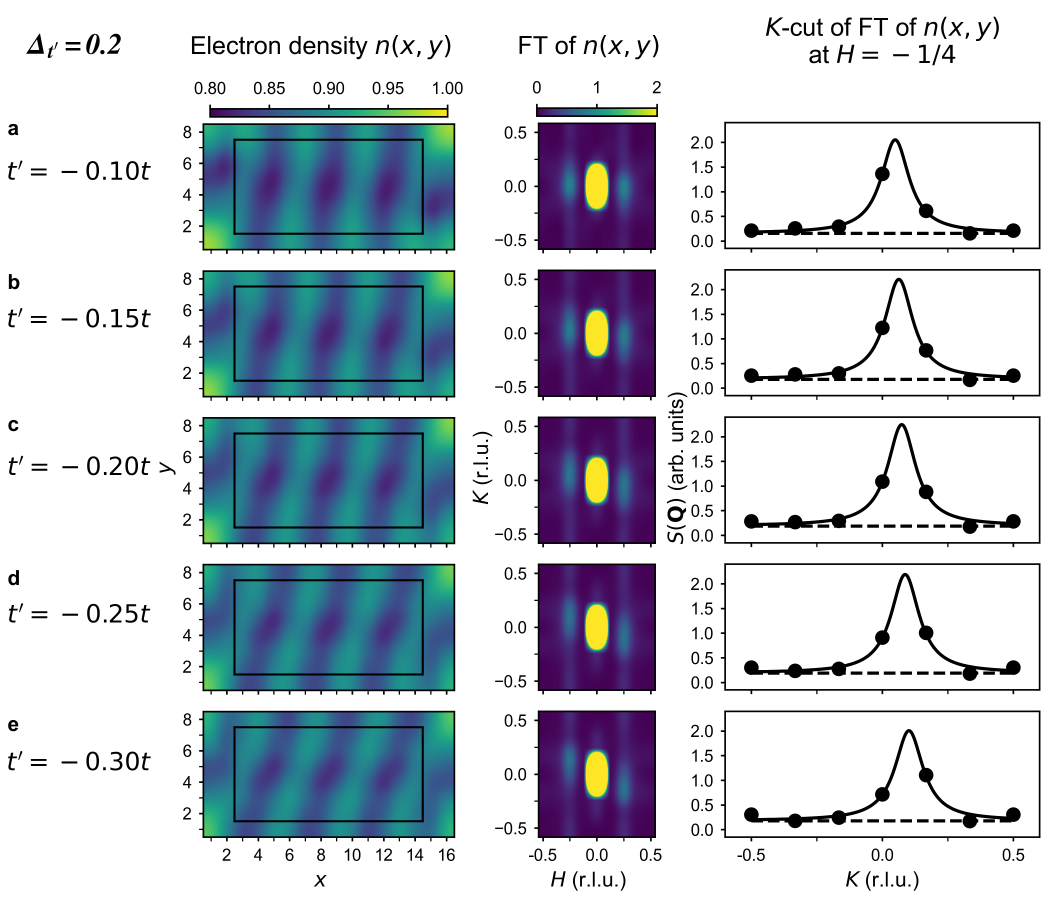}%
 \caption{\textbf{Charge stripes for $\Delta_{t'}=0.2$ at various $t'$ values calculated using the \gls*{DMRG} method}. This includes the electron density distributions $n(x,y)$ (left column), the corresponding Fourier transform (middle column) and $K$-cuts of the charge ordering peaks at $H=-0.25$ (right column). The rectangles in the left column depict regions used in the Fourier transform. The solid lines in the right column are Lorentzian fits with a constant background (dashed lines). The maps in the left and middle columns are produced using multiquadric interpolation method \cite{matplotlib}.}
 \label{fig:charge2}
\end{figure}

\begin{figure}[htbp]
 \includegraphics[scale=1]{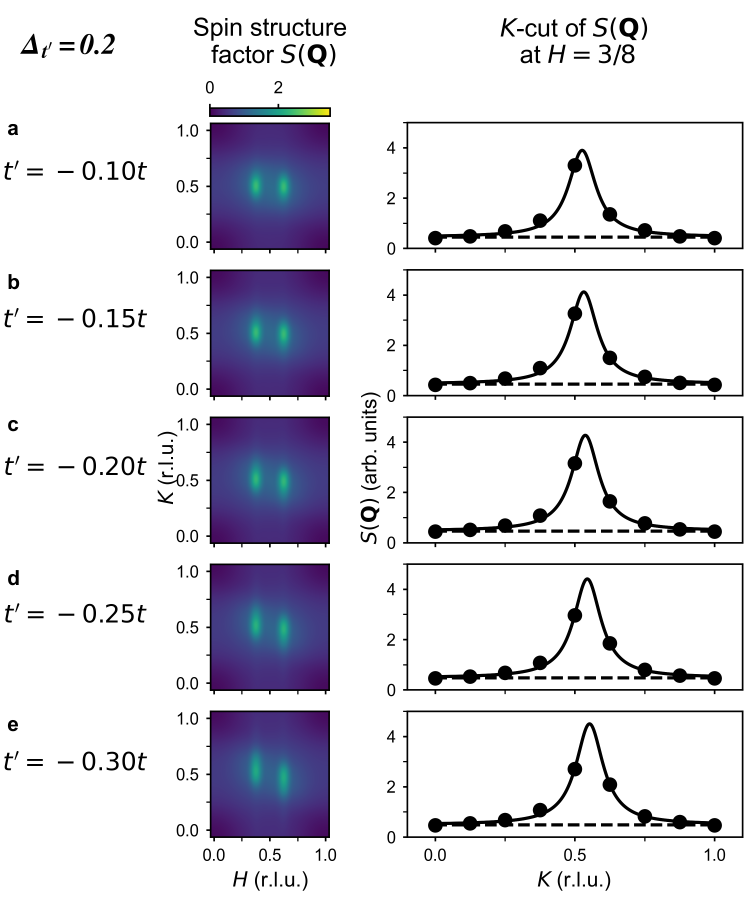}%
 \caption{\textbf{Spin stripes for $\Delta_{t'}=0.2$ at various $t'$ values calculated using the \gls*{DMRG} method}. This includes the static spin structure factors $S(\mathbf{Q})$ (left column) and $K$-cuts of the spin ordering peaks in $S(\mathbf{Q}$) at $H=0.375$ (right column). The solid lines in the right column are Lorentzian fits with a constant background (dashed lines). The maps in the left column are produced using multiquadric interpolation method \cite{matplotlib}.}
 \label{fig:spin2}
\end{figure}

\begin{figure}[htbp]
 \includegraphics[scale=1]{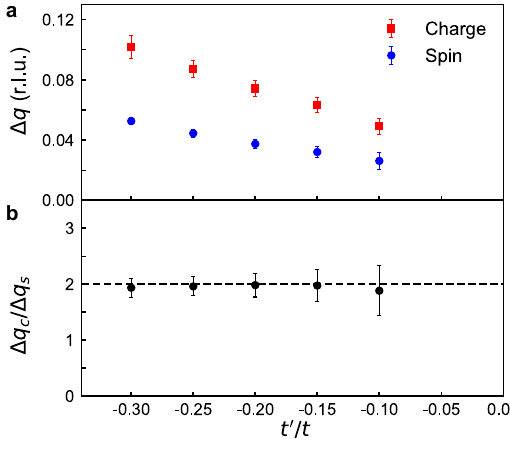}%
 \caption{\textbf{\textit{Y}-shift for both charge and spin ordering peaks for $\Delta_{t'}=0.2$ from \gls*{DMRG} calculations and their ratios at various $t'$ values}. $\Delta q_c$ ($\Delta q_s$) is the fitted charge (spin) peak positions along $K$ direction shown in the right column of Supplementary Fig.~\ref{fig:charge2} (\ref{fig:spin2}) relative to the position if $\Delta_{t'}=0$, i.e., $\Delta q_c=|q_c-0|$ ($\Delta q_s=|q_s-0.5|$). Error bars correspond to $\pm \sigma$, where $\sigma$ is the standard deviation.
 }
 \label{fig:fit2}
\end{figure}

\begin{figure}[htbp]
 \includegraphics[scale=1]{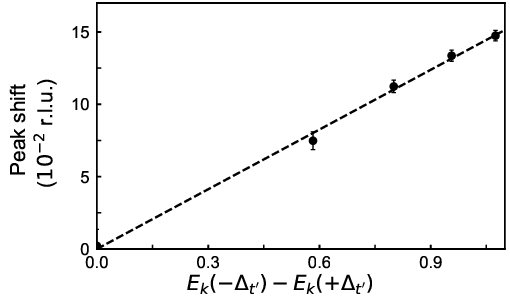}%
 \caption{\textbf{\textit{Y}-shift as a function of the kinetic energy difference between the $(1-\Delta_{t'})$-bonds and the $(1+\Delta_{t'})$-bonds}. \textit{Y}-shift is measured by the peak shift shown in Fig.~6. The dashed line is a linear fit to the data. Error bars correspond to $\pm \sigma$, where $\sigma$ is the standard deviation.
 }
 \label{fig:YshiftvsEnergydiff}
\end{figure}

\begin{figure}[htbp]
 \includegraphics[angle=90,height=0.95\textheight]{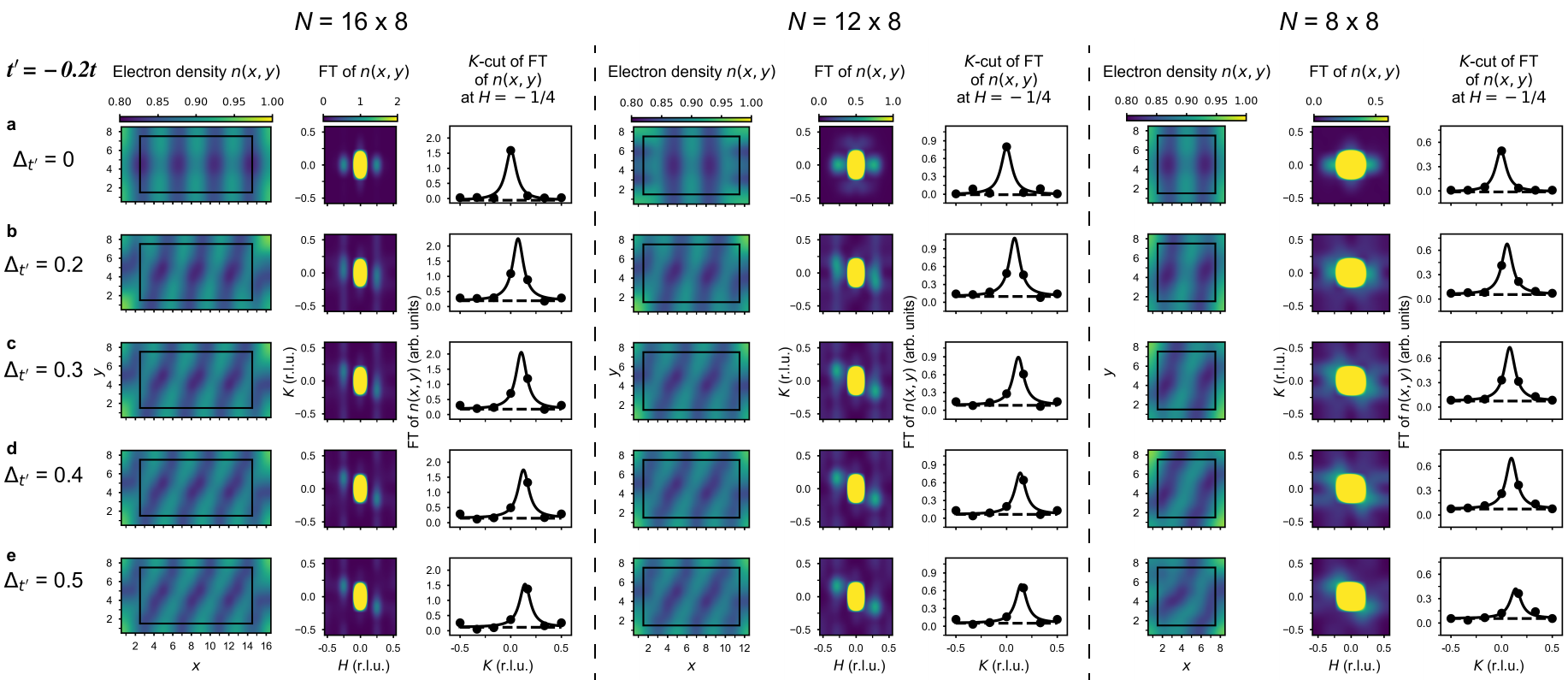}
 \captionstyle{\centering}
 \caption{\label{fig:charge_comparision}(Continued on the following page.)}
\end{figure}
\begin{figure}[htbp]
 \contcaption{\textbf{Comparison of charge stripes calculated using the \gls*{DMRG} method with different system sizes}. From left to right, the system sizes are $N=16\times 8$, $N=12\times 8$, and $N=8\times 8$, respectively. For each system size, the column contains three sub-columns displaying the electron density distributions $n(x,y)$, the corresponding Fourier transform, and $K$-cuts of the charge ordering peaks at $H=-0.25$. The same $t'$ value ($t'=-0.2t$) and hole doping concentration ($\delta=12.5\%$) are used in the calculations. The results for $N=16\times 8$ use the same data shown in Supplementary Fig.~\ref{fig:charge1}. The rectangles in the left sub-columns depict regions used in the Fourier transform. The solid lines in the right sub-columns are Lorentzian fits with a constant background (dashed lines). The maps in the left and middle sub-columns are produced using multiquadric interpolation method \cite{matplotlib}.
 }
\end{figure}

\begin{figure}[htbp]
 \includegraphics[angle=90,height=0.95\textheight]{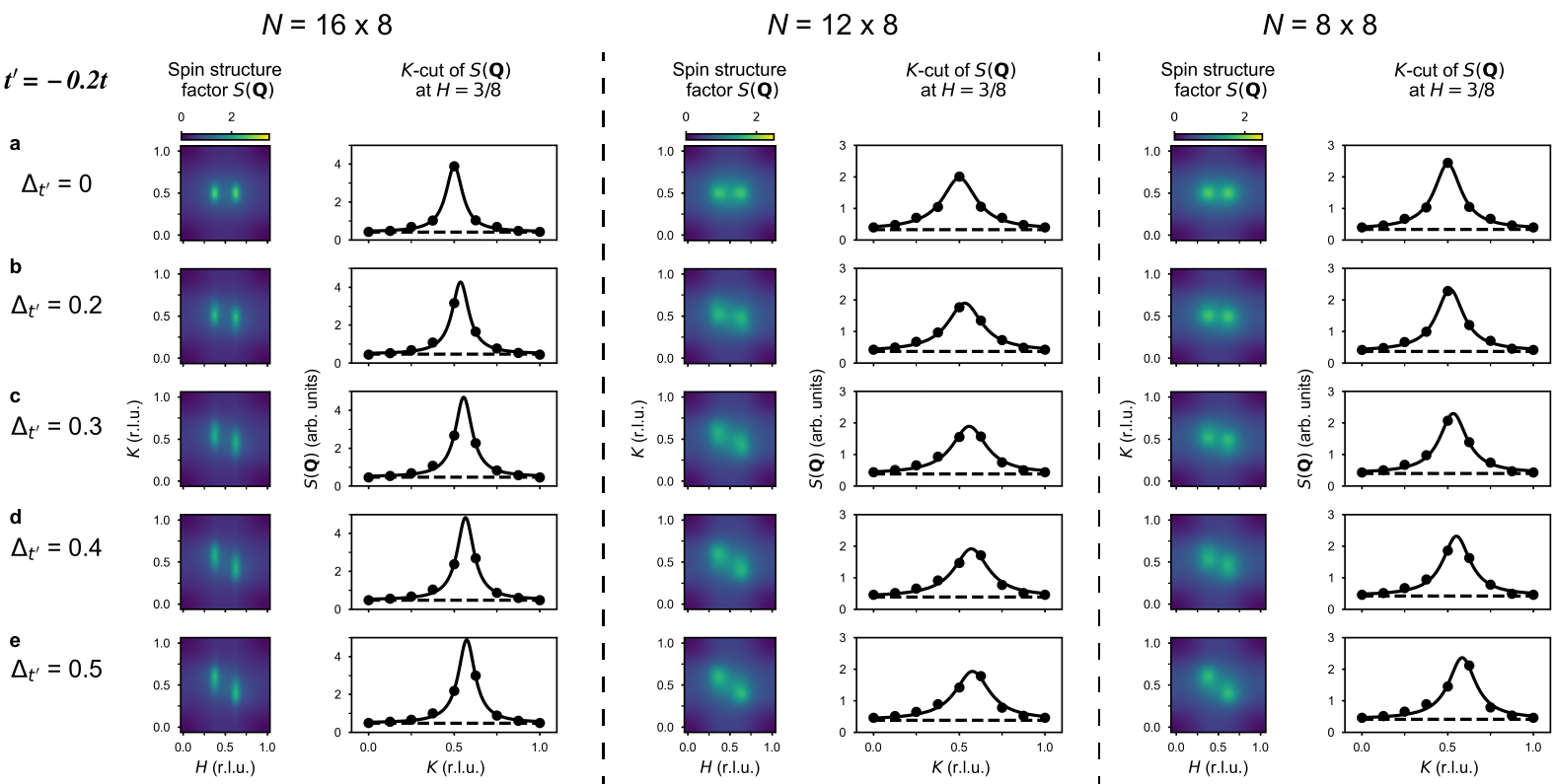}
 \captionstyle{\centering}
 \caption{\label{fig:spin_comparision}(Continued on the following page.)}
\end{figure}
\begin{figure}[htbp]
 \contcaption{\textbf{Comparison of spin stripes calculated using the \gls*{DMRG} method with different system sizes}. From left to right, the system sizes are $N=16\times 8$, $N=12\times 8$, and $N=8\times 8$, respectively. For each system size, the column contains two sub-columns displaying the static spin structure factors $S(\mathbf{Q})$ and $K$-cuts of the spin ordering peaks in $S(\mathbf{Q}$) at $H=0.375$. The same $t'$ value ($t'=-0.2t$) and hole doping concentration ($\delta=12.5\%$) are used in the calculations. The results for $N=16\times 8$ use the same data shown in Supplementary Fig.~\ref{fig:spin1}. The solid lines in the right sub-columns are Lorentzian fits with a constant background (dashed lines). The maps in the left sub-columns are produced using multiquadric interpolation method \cite{matplotlib}.
 }
\end{figure}

\begin{figure}[htbp]
 \includegraphics[scale=0.9]{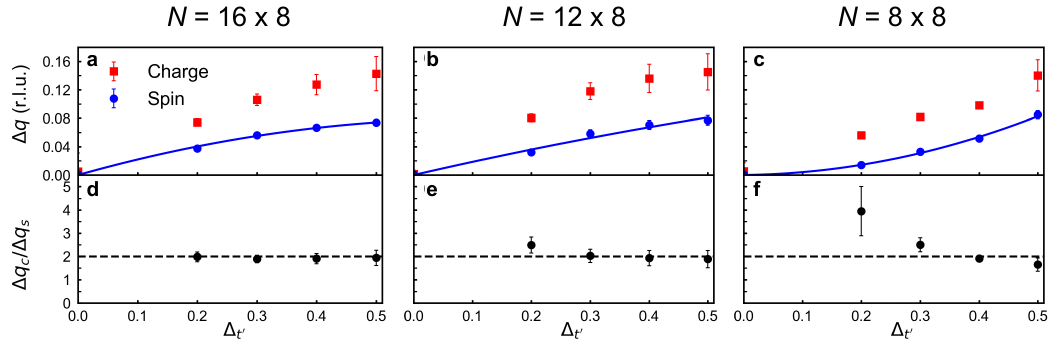}%
 \caption{\textbf{Comparison of peak shifts from the \gls*{DMRG} calculations with different system sizes}. From left to right, the system sizes are $N=16\times 8$, $N=12\times 8$, and $N=8\times 8$, respectively. \textbf{a}--\textbf{c} show the peak shifts for both charge and spin ordering peaks, while \textbf{d}--\textbf{f} show their ratios. $\Delta q_c$ ($\Delta q_s$) is the fitted charge (spin) peak positions along $K$ direction shown in the right sub-columns of Supplementary Fig.~\ref{fig:charge_comparision} (\ref{fig:spin_comparision}) relative to the position if $\Delta_{t'}=0$, i.e., $\Delta q_c=|q_c-0|$ ($\Delta q_s=|q_s-0.5|$). The results for $N=16\times 8$ use the same data shown in Supplementary Fig.~\ref{fig:fit1}. The solid blue lines in \textbf{a}--\textbf{c} are quadratic fits to the data. Error bars correspond to $\pm \sigma$, where $\sigma$ is the standard deviation.
 }
 \label{fig:fit_comparision1}
\end{figure}

\begin{figure}[htbp]
 \includegraphics[scale=1]{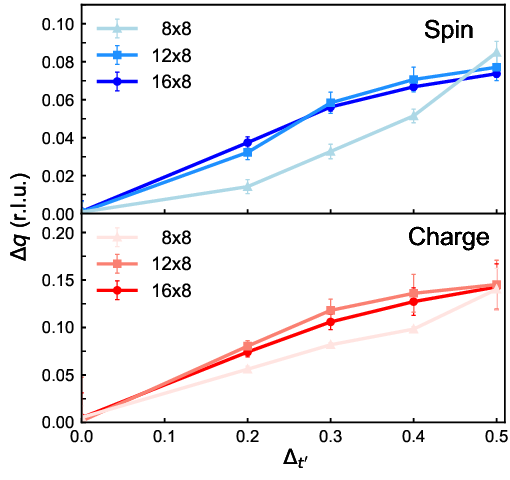}%
 \caption{\textbf{Another comparison of peak shifts from the \gls*{DMRG} calculations with different system sizes}. The data are the same as those in Supplementary Figs.~\ref{fig:fit_comparision1}. Error bars correspond to $\pm \sigma$, where $\sigma$ is the standard deviation.
 }
 \label{fig:fit_comparision2}
\end{figure}

\clearpage

\bibliographystyle{naturemag}
\bibliography{ref}